\documentclass[12pt]{article}
\usepackage{amssymb}
\usepackage{amsmath}
\usepackage{epsfig}
\usepackage{psfrag}
\usepackage{latexsym}
\usepackage{euscript}
\usepackage{multirow}
\usepackage{bbold}
\usepackage{cite}

\usepackage{rotating}
\usepackage{verbatim}
\usepackage{array}
\usepackage{cancel}
\usepackage{arydshln}
\def\be{\begin{equation}}
\def\ee{\end{equation}}

\def\bea{\begin{eqnarray}}
\def\eea{\end{eqnarray}}
\def\nn{\nonumber}

\def\be{\begin{equation}}
\def\ee{\end{equation}}
\def\bc{\begin{center}}
\def\ec{\end{center}}
\def\bea{\begin{eqnarray}}
\def\eea{\end{eqnarray}}

\def\nn{\nonumber}

\catcode`@=11
\def\marginnote#1{}
\newcount\hour
\newcount\minute
\newtoks\amorpm
\hour=\time\divide\hour by60
\minute=\time{\multiply\hour by60 \global\advance\minute by-\hour}
\edef\standardtime{{\ifnum\hour<12 \global\amorpm={am}%
        \else\global\amorpm={pm}\advance\hour by-12 \fi
        \ifnum\hour=0 \hour=12 \fi
        \number\hour:\ifnum\minute<10 0\fi\number\minute\the\amorpm}}
\edef\militarytime{\number\hour:\ifnum\minute<10 0\fi\number\minute}
\def\draftlabel#1{{\@bsphack\if@filesw {\let\thepage\relax
   \xdef\@gtempa{\write\@auxout{\string
      \newlabel{#1}{{\@currentlabel}{\thepage}}}}}\@gtempa
   \if@nobreak \ifvmode\nobreak\fi\fi\fi\@esphack}
        \gdef\@eqnlabel{#1}}
\def\@eqnlabel{}
\def\@vacuum{}
\def\draftmarginnote#1{\marginpar{\raggedright\scriptsize\tt#1}}
\def\draft{\oddsidemargin 0.0truein
        \def\@oddfoot{\sl preliminary draft \hfil
        \rm\thepage\hfil\sl\today\quad\militarytime}
        \let\@evenfoot\@oddfoot \overfullrule 3pt
        \let\label=\draftlabel
        \let\marginnote=\draftmarginnote
   \def\@eqnnum{(\theequation)\rlap{\kern\marginparsep\tt\@eqnlabel}%
\global\let\@eqnlabel\@vacuum}  }
\catcode`@=12

\begin{document}
\begin{titlepage}
\vspace*{-1cm}
\phantom{hep-ph/***}
\hfill{DFPD-11/TH/18}\\
\vskip 2.5cm
\begin{center}
\mathversion{bold}
{\Large\bf Finite Modular Groups and Lepton Mixing}
\mathversion{normal}
\end{center}
\vskip 0.5  cm
\begin{center}
{\large Reinier de Adelhart Toorop}~\footnote{e-mail address: reintoorop@nikhef.nl}
\\
\vskip .2cm
Nikhef Theory Group
\\
Science Park 105, 1098 XG Amsterdam, The Netherlands
\\
\vskip .2cm
and
\\
{\large Ferruccio Feruglio}~\footnote{e-mail address: ferruccio.feruglio@pd.infn.it},
{\large Claudia Hagedorn}~\footnote{e-mail address: claudia.hagedorn@pd.infn.it}
\\
\vskip .2cm
Dipartimento di Fisica e Astronomia `G.~Galilei', Universit\`a di Padova 
\\
INFN, Sezione di Padova, Via Marzolo~8, I-35131 Padua, Italy
\end{center}
\vskip 0.7cm
\begin{abstract}
\noindent We study lepton mixing patterns which are derived from finite modular groups $\Gamma_N$, 
requiring subgroups $G_\nu$ and $G_e$ to be preserved in the neutrino and charged lepton sectors, respectively. We show that
only six groups $\Gamma_N$ with $N=3,4,5,7,8,16$ are relevant. A comprehensive analysis is presented for 
$G_e$ arbitrary and $G_\nu=Z_2\times Z_2$,  as demanded if neutrinos are Majorana particles. We discuss interesting 
patterns arising from both groups $G_e$ and $G_\nu$ being arbitrary. Several of the most promising patterns are 
specific deviations from tri-bimaximal mixing, all predicting $\theta_{13}$ non-zero as favoured by the latest experimental data. 
We also comment on prospects to extend this idea to the quark sector. 
\end{abstract}
\end{titlepage}
\setcounter{footnote}{0}
\vskip2truecm

\section{Introduction}
\label{sec:intro}

The origin of fermion mixing is one of the most fascinating unsolved problems of particle physics and 
the unexpected difference between quark and lepton mixing 
is very puzzling. The Cabibbo-Kobayashi-Maskawa (CKM) mixing matrix $V_{CKM}$ is numerically close to the identity matrix \cite{pdg}, i.e. up and down quark mass matrices are only 
slightly misaligned in flavour space. On the contrary,  all entries of the Pontecorvo-Maki-Nakagawa-Sakata  (PMNS) mixing matrix $U_{PMNS}$
are of order one, with the exception of $U_{e3}$, see \cite{T2K,MINOS,DC} and \cite{fogli,schwetz,maltoni}, indicating a substantial misalignment of charged lepton and neutrino mass matrices.
The lightness of neutrinos suggests their Majorana character, an important feature still waiting 
for experimental confirmation. This feature might play a key role in explaining the difference between quarks and leptons.
Searching for an explanation of the observed fermion mixing patterns it is plausible to consider a limit  in which such patterns become simple. 
For instance, $V_{CKM}$ equals the identity matrix in a rough approximation, 
a feature which can arise from a symmetry or dynamical property the theory possesses in a certain limit.
If small corrections to such a limit are taken into account, the experimental data can be accommodated.

Several simple patterns for the PMNS matrix have been
determined and a possible explanation of their origin has been 
formulated in terms of discrete flavour symmetries present in the underlying theory, for reviews see \cite{reviews}.
As has been discussed in detail in the literature \cite{Lam0708,BHL07}, a 
 framework in which the misalignment between neutrino and charged lepton mass matrices is 
associated with the non-trivial breaking of a flavour symmetry is particularly interesting and predictive. The idea is as follows:
the underlying theory is invariant under a  flavour group $G_f$ 
while, at the leading order (LO), the neutrino and the charged lepton sectors
are separately invariant under two different subgroups of $G_f$: $G_\nu$ and $G_e$, respectively.
\footnote{The intersection of $G_\nu$ and $G_e$ is empty, because no non-trivial flavour group is compatible with
low energy data.}
If left-handed leptons transform as a three-dimensional irreducible representation under $G_f$, 
the lepton mixing can be fixed by the groups $G_\nu$, $G_e$ and their relative embedding into $G_f$. 
In particular, the LO prediction for the mixing angles is independent of the parameters of the 
theory.\footnote{It has been shown in \cite{HCHM_fl} that (holographic) composite Higgs models, in particular their five-dimensional realizations, are a suitable framework
for this kind of approach.}

If there are three generations of neutrinos and these are Majorana particles, 
the maximal invariance group of the neutrino mass matrix is a Klein group $Z_2 \times Z_2$ \cite{Lam0708,Lam08}.
We thus set $G_\nu=Z_2\times Z_2$ and we choose $G_e$ to be a discrete group as well. In this case also the group $G_l$ generated by $G_\nu$ and $G_e$ is discrete.
Furthermore, we require that $G_l$ is finite, because in the case of $G_l$ infinite we are eventually able to accommodate any mixing pattern for leptons and 
thus this approach loses its predictive power.
Usually $G_l$ does not coincide with $G_f$, because additional factors are present in $G_f$ in order to reduce
the number of operators allowed at LO,
or to extend the construction to the quark sector. However, the lepton mixing, resulting at LO,
is determined by $G_l$ and its subgroups $G_\nu$ and $G_e$. 

Interesting mixing patterns
arise from discrete groups $G_l$ such as  $S_4$ 
and  $A_5$. 
For example, it is possible to generate patterns in which the
atmospheric mixing angle $\theta_{23}$ is maximal and the reactor mixing angle $\theta_{13}$ vanishes. They
only differ in the value predicted for the solar mixing angle:
$\sin^2\theta_{12}=1/3$ for tri-bimaximal mixing \cite{HPS}, $\sin^2\theta_{12}=1/2$ for bimaximal
mixing \cite{BMorg} and $\tan\theta_{12}=1/\phi$ with $\phi=(1+\sqrt{5})/2$ for the so-called golden ratio mixing \cite{GRpro}.
The first two are derived from $S_4$, see \cite{Lam11} and earlier \cite{S4TB,S4BM},\footnote{We do not include the example of tri-bimaximal mixing derived in the context of 
$A_4$ models \cite{A4examples} because in this case the Klein group preserved in the neutrino sector is partially accidental.}
while the latter is realized in models employing $A_5$ as flavour symmetry \cite{GR}.
Small corrections to these patterns can lead to a good agreement with the present data.

The scenario outlined is not the most general one: in concrete models
$G_\nu$ and $G_e$ might be partly or entirely accidental with no link to the underlying flavour symmetry $G_f$, see e.g. \cite{A4examples,King_indirect}.
Moreover, at LO the mixing angles are determined from group theory alone and thus no direct relation between masses and mixing angles is obtained, 
contrary to what might be expected in the quark sector, see e.g. \cite{GST}. This is one of the reasons why extensions to the quark sector are believed not to be straightforward. 
Nevertheless, it is important to explore the predictive power and advantages of such an idea.

An open question is whether new interesting mixing patterns for leptons can arise 
by extending the list of possible groups $G_l$.
In this paper we provide a partial answer by analyzing the infinite sequence of groups $\Gamma_N$ to which $A_4$, $S_4$ and $A_5$ belong.
Indeed, these groups can be generated by two elements $S$ and $T$ satisfying the relations $S^2=(ST)^3=T^N=E$ with 
$N=3,4,5$ for $A_4$, $S_4$ and $A_5$, respectively ($E$ denotes the neutral element of the group). The naive extension of this sequence to the case $N>5$ leads to infinite discrete groups \cite{CoxeterMoser}, 
that are less attractive from the model building point of view and that we exclude from our analysis. A restriction to finite groups can be achieved by 
considering the so-called finite modular groups $\Gamma_N$ ($N>1$ integer), subgroups of the inhomogeneous modular group $\Gamma$.
The groups $A_4$, $S_4$ and $A_5$ are isomorphic to $\Gamma_N$ for $N=3,4,5$ respectively.
By choosing $G_l$ as $\Gamma_N$ (or as an appropriate subgroup of $\Gamma_N$), 
we identify all associated lepton mixing patterns. 
The requirement that the group possesses three-dimensional irreducible representations, needed to accommodate
left-handed lepton doublets, reduces the number of cases considerably and a comprehensive analysis can be performed 
by considering the six groups corresponding to $N=3,4,5,7,8,16$.  In doing so, 
we determine the subgroups $Z_2\times Z_2$, all relevant candidates for $G_e$
and the resulting mixing patterns.  In a separate publication \cite{dATFH11} we have already presented two specific mixing patterns that are of particular interest in the light of the recent results of the 
T2K \cite{T2K}, MINOS \cite{MINOS} and Double Chooz \cite{DC} collaborations, because they predict $\theta_{13} \sim 0.1 \div 0.2$.
After having studied all cases for $G_\nu=Z_2 \times Z_2$, we also discuss promising patterns for any possible choice of $G_\nu$ and $G_e$
and we comment on the applicability of this idea to the quark sector. 

The structure of the paper is as follows: in section 2 we summarize important results about finite modular groups and their representations. In particular we show 
that the groups relevant for our discussion are  $\Gamma_N$ for $N=3,4,5,7,8,16$. In section 3 we first state our
assumptions and then we derive for each group the mixing patterns arising from $G_\nu=Z_2\times Z_2$ and any choice of $G_e$.
We present some interesting patterns obtained by relaxing the assumption that neutrinos are Majorana particles in section 4 and
we comment on the possible extension of our analysis to the quark sector in section 5.
Finally, we conclude in section 6. Two appendices are added containing further details on the groups.

\section{Finite modular groups and their representations}
\label{sec:modular}

As mentioned in the introduction, flavour symmetries containing the groups $S_4$ and $A_5$ 
(and, taking into account also accidental symmetries, $A_4$)
can give rise to interesting lepton mixing patterns. 
They can be cast into a common presentation given in terms of two generators
$S$ and $T$ satisfying:
\be 
S^2=E~~,~~(ST)^3=E~~,~~~T^N=E~~~,
\label{pres}
\ee
with $N=4,5$ ($N=3$), respectively, and $E$ being the neutral element of the group.  An obvious question is whether this presentation extends to other
finite groups for $N>5$.  Without any additional requirement on $S$ and $T$, the relations in eq.(\ref{pres})
define in general an infinite group for $N > 5$ \cite{CoxeterMoser}. Though not excluded as candidates for a flavour symmetry, infinite discrete groups 
have the disadvantage that they can possess infinitely many irreducible representations of a given dimensionality,
which makes them less appealing for model building. This fact is related to the argument given in the introduction that admitting infinite groups eventually allows to reproduce any mixing
pattern. Thus, in order to extend the list of $A_4$, $S_4$ and $A_5$ we follow
a different approach: we note that these three groups are members of the infinite sequence of finite modular groups $\Gamma_N$ \cite{schoen} and thus consider the latter in our study. 
One important requirement in our scenario is that the flavour group possesses three-dimensional irreducible representations
to which the left-handed lepton doublets are assigned.
In this section we thus discuss which finite modular group satisfies the latter condition and show that it is sufficient to consider only six groups $\Gamma_N$ with $N=3,4,5,7,8,16$.

\subsection{Finite modular groups}

The inhomogeneous modular group $\Gamma$ is the group of linear fractional transformations acting on a complex variable $z$:
\footnote{We follow the notation used in \cite{gun}. Other authors \cite{schoen} call instead the homogeneous modular group $\Gamma$.}
\be
z\to \frac{a z+b}{c z+d}
\ee
with $a$, $b$, $c$ and $d$ being integers and $a d-b c=1$. Obviously, a transformation characterized by parameters \{$a$, $b$, $c$, $d$\} is identical to the one defined by \{-$a$, -$b$, -$c$, -$d$\}. 
As consequence, $\Gamma$ is isomorphic to $PSL(2,Z)=SL(2,Z)/\{\pm \mathbb{1}\}$ where  $SL(2,Z)$ is the group of two-by-two matrices with integer entries and determinant equal to one.
The group $\Gamma$ can be generated with two elements $S$ and $T$ which fulfill
\cite{gun}:
\be
S^2=E~~,~~(ST)^3=E~~
\label{modular}
\ee
and are given by the transformations
\be
S: z \rightarrow -\frac{1}{z}~~~,~~~~~~~~~~
T: z \rightarrow z+1~~~.
\ee
They can be represented by  the following two matrices of $SL(2,Z)$:
\be
S=\left(
\begin{array}{cc}
0&1\\
-1&0
\end{array}
\right)~~~,~~~~~~~~~~
T=\left(
\begin{array}{cc}
1&1\\
0&1
\end{array}
\right)~~~.
\label{sandt}
\ee

We can generalize these groups by replacing integers with integers modulo $N$ \cite{schoen}. For a natural number $N>1$ the group $SL(2,Z_N)$ is defined as the group of two-by-two matrices with entries that are integers modulo $N$ and determinant equal to one modulo $N$. Then the inhomogeneous finite modular groups
\footnote{Note that the notation in the literature varies in this case \cite{gun,schoen}.}
\be
\Gamma_N = SL(2,Z_N)/\{\pm \mathbb{1}\}
\ee
are defined by identifying matrices in $SL(2,Z_N)$ which are related by an overall sign. For each $N$ these groups are finite.
The order of $SL(2,Z_N)$ is \cite{gun,schoen}
\be
\Big\vert SL(2,Z_N) \Big\vert= N^3 \prod_{p\vert N}\left(1-\frac{1}{p^2}\right)
\label{orderSL2ZN}
\ee
with the product ranging over the prime divisors $p$ of $N$.
For $N=2$, the matrices $\mathbb{1}$ and $-\mathbb{1}$
are indistinguishable and therefore $\Gamma_2$ and $SL(2,Z_2)$ are isomorphic: $\Gamma_2\simeq SL(2,Z_2)$. For $N>2$ they are distinguishable and the order of $\Gamma_N$ is
\be
\Big\vert \Gamma_N \Big\vert=\frac{1}{2} \Big\vert SL(2,Z_N)\Big\vert~~~~\mbox{for}~~~~~~~N>2~~~.
\ee
In table 1 we list the orders of $\Gamma_N$ for $2\le N\le 16$.
\\[0.2cm]
\begin{table}[h]
\begin{center}
\begin{tabular}{|c|c|c|c|c|c|c|c|c|c|c|c|c|c|c|c|}
\hline
$N$&2&3&4&5&6&7&8&9&10&11&12&13&14&15&16\\
\hline
$|\Gamma_N|$&6&12&24&60&72&168& 192 & 324 & 360 & 660&576 &1092 &1008 &1440 &1536 \\
\hline
\end{tabular}
\end{center}
\caption{\label{order} Order of $\Gamma_N$ for $2\le N\le 16$.}
\end{table}
\vspace{0.2cm}

We investigate the series $\Gamma_N$ in the following. First of all, note that in general 
\be
T^N=E
\label{tn}
\ee
is fulfilled  in the groups $\Gamma_N$. The smallest group $\Gamma_2$ is isomorphic to $S_3$. For $N=3,4,5$ the isomorphisms $\Gamma_3\simeq A_4$, $\Gamma_4\simeq S_4$ 
and $\Gamma_5\simeq A_5$ hold  \cite{schoen}.
For $N$ larger than five, however, the relations in eqs. (\ref{modular}) and (\ref{tn}) are not sufficient in order to render the group $\Gamma_N$ finite.
We specify the additional relations after having determined the groups relevant for our analysis.

\mathversion{bold}
\subsection{Irreducible representations of $SL(2,Z_N)$}
\mathversion{normal}

In this subsection we discuss the representations of the groups $SL(2,Z_N)$ \cite{nobs,eholzer}.
Since it exists an homomorphism between $\Gamma_N$ and $SL(2,Z_N)$ \cite{schoen}, which is an isomorphism for $N=2$, 
all representations of $\Gamma_N$ are also representations of $SL(2,Z_N)$.
We first recall the classification of the irreducible representations of $SL(2,Z_N)$. In this way we obtain
all representations of the group $\Gamma_N$ we are interested in, and additional representations that are eventually discarded.
We distinguish the three possible cases for $N$ ($\lambda, \lambda_p \in \mathbb{N}$)
\begin{itemize}
\item[1)] $N$ is prime.
\item[2)] $N=p^\lambda$ with $p$ prime and $\lambda>1$.
\item[3)] $N=\prod \limits_p p^{\lambda_p}$ with $p$ prime and $\lambda_p\ge 1$.
\end{itemize}

We start with case 1). As remarked before, if $N=2$, we have $\Gamma_2 \simeq S_3$, which has two one-dimensional and one two-dimensional representations, but no
irreducible three-dimensional ones.
The dimensions $d$ and multiplicities $\mu$ of the irreducible representations  of $SL(2, Z_p)$ where $p$ is an \textit{odd} prime are given in table \ref{tab_dim_prime}.
\\[0.2cm]
\begin{table}[h]
\begin{center}
\begin{tabular}{|c|c|c|c|c|c|c|}
\hline
$d$&$1$&$p+1$&$p-1$&$\frac{1}{2}(p+1)$&$\frac{1}{2}(p-1)$&$p$\\
\hline
$\mu$&1&$\frac{1}{2}(p-3)$&$\frac{1}{2}(p-1)$&$2$&$2$&$1$\\
\hline
\end{tabular}
\end{center}
\caption{\label{tab_dim_prime} Dimensions $d$ and multiplicities $\mu$ of the irreducible representations of $SL(2,Z_p)$, $p$ being an odd prime \cite{nobs,eholzer}.}
\end{table}
\vspace{0.2cm}

One can easily check that
\be
\sum_i d_i^2 \mu_i=p^3\left(1-\frac{1}{p^2}\right)
\ee
gives the correct order of the group. We find that $SL(2, Z_p)$ has three-dimensional irreducible representations only for $p=3, 5$ and 7: one for $SL(2,Z_3)$
and two for both $SL(2,Z_5)$ and $SL(2,Z_7)$, see table \ref{tab_dim_prime}.

Next, we consider the case in which $N$ is a power of a prime, i.e. $N=p^\lambda$ . We separately discuss the following two cases:  
$p$ is an odd prime and $p=2$. In table \ref{tab_dim_power_prime} we list the irreducible representations $d$ of $SL(2, Z_{p^\lambda})$ with $p>2$ and $\lambda>1$, and the multiplicities $\mu$ of these representations.
This table has to be read as follows: given an integer $\bar{\lambda}>1$, all groups $SL(2,Z_{p^\lambda})$ with $\lambda<\bar{\lambda}$ are
homomorphic to the group $SL(2,Z_{p^{\bar{\lambda}}})$ \cite{nobs,eholzer}. It follows that the representations of $SL(2,Z_{p^\lambda})$ with $1\le \lambda< \bar{\lambda}$
are also representations of the group $SL(2,Z_{p^{\bar{\lambda}}})$. The irreducible representations of $SL(2,Z_{p^{\bar{\lambda}}})$ are given by those
of table \ref{tab_dim_prime} ($\lambda=1$) and by those listed in table \ref{tab_dim_power_prime}, with $\lambda=2,...,\bar{\lambda}$. For instance, if $\bar{\lambda}=3$ both possibilities $\lambda=2$ and $\lambda=3$ have to be considered when using table  \ref{tab_dim_power_prime}. Again, it can be checked that 
the order of $SL(2,Z_{p^{\bar{\lambda}}})$, using tables \ref{tab_dim_prime} and \ref{tab_dim_power_prime}, is $p^{3 \bar{\lambda}}(1-1/p^2)$ in agreement with eq. (\ref{orderSL2ZN}).
With the help of table \ref{tab_dim_power_prime} it is simple to prove that for $p>2$ and $\lambda>1$ there are no other three-dimensional irreducible representations,
apart from those already given in table \ref{tab_dim_prime}. This concludes the discussion for $p>2$.

\begin{table}[h!]
\begin{center}
\begin{tabular}{|c|c|c|c|}
\hline
$d$&$p^{\lambda-1}(p+1)$&$p^{\lambda-1}(p-1)$&$\frac{1}{2}p^{\lambda-2}(p^2-1)$\\
\hline
$\mu$&$\frac{1}{2}p^{\lambda-2}(p-1)^2$&$\frac{1}{2}p^{\lambda-2}(p^2-1)$&$4 p^{\lambda-1}$\\
\hline
\end{tabular}
\begin{minipage}{13cm}
\caption{\label{tab_dim_power_prime} Dimensions $d$ and multiplicities $\mu$ of the additional irreducible representations of $SL(2,Z_{p^\lambda})$, $p$ being an odd prime and $\lambda>1$  \cite{nobs,eholzer}.
See text for explanations.}
\end{minipage}
\end{center}
\end{table}

The case $p=2$ is more complicated and a separate discussion for each small $\lambda$ is needed. Similar to the former case, also here the representations of $SL(2,Z_{2^\lambda})$ are representations
of the group $SL(2,Z_{2^{\bar{\lambda}}})$ for $1\le \lambda< \bar{\lambda}$. For $\lambda>4$ there are no three-dimensional irreducible representations, different from those
already induced by $\lambda=2,3,4$ \cite{nobs,eholzer}. In table \ref{tab_dim_power2} we summarize the irreducible representations of $SL(2,Z_{2^\lambda})$ and their multiplicities for $\lambda=1,2,3,4$. We conclude that it is sufficient to analyze $\Gamma_4$, $\Gamma_8$ and $\Gamma_{16}$ in order to cover all cases which are relevant 
from a model building point of view.
\\[0.2cm]
\begin{table}[h!]
\begin{center}
\begin{tabular}{|c|c|c|c|c|c|c|c|c||c|}
\hline
$d$&1&2&3&4&6&8&12&24&{\tt order}\\
\hline
$SL(2,Z_2)$&2&1&&&&&&&6\\
\hline
$SL(2,Z_4)$&4&2&4&&&&&&48\\
\hline
$SL(2,Z_8)$&4&6&12&2&6&&&&384\\
\hline
$SL(2,Z_{16})$&4&6&28&2&26&6&2&2&3072\\
\hline
\end{tabular}
\end{center}
\caption{\label{tab_dim_power2} Dimensions $d$ and multiplicities of the irreducible representations of $SL(2,Z_{2^\lambda})$, for $\lambda<5$. For each group
all the irreducible representations and the order of the group are listed \cite{nobs,eholzer}.}
\end{table}
\vspace{0.2cm}

Lastly, we consider the case in which $N$ is a product of primes
\be
\label{Nprod}
N=\prod_p p^{\lambda_p}~~~~~~~,~~~\lambda_p\ge1~~~.
\ee
The group $SL(2,Z_N)$ factorizes
as
\be
SL(2,Z_N)=\prod_p SL(2,Z_{p^{\lambda_p}}) \, .
\ee
Since the three-dimensional representations of these product groups are constructed by using the three-dimensional representations of one of the groups and one-dimensional representations of all the others
\cite{eholzer}, the cases in which $N$ is a product of the form given in eq.(\ref{Nprod})  cannot give rise to independent three-dimensional representations.

In conclusion, all independent three-dimensional representations of the finite modular groups $\Gamma_N$ can be studied by considering the six groups $SL(2,Z_N)$ $(N=3,4,5,7,8,16)$.
We find 33 distinct irreducible triplets. From table \ref{tab_dim_prime} we see that one is associated with the case $N=3$, two are related to the case $N=5$ and two to $N=7$. Moreover,
using table \ref{tab_dim_power2}, we count four irreducible triplets corresponding to $N=4$, while in the case of $N=8$ eight additional irreducible triplets are encountered and another 16 independent irreducible triplets are associated with $N=16$. 

\mathversion{bold}
\subsection{Three-dimensional irreducible representations of $\Gamma_N$}
\mathversion{normal}

The 33 triplets of the groups $SL(2,Z_N)$ ($N=3,4,5,7,8,16$) do not all fulfill the relations given in eq.(\ref{modular}) and thus not all of them are also representations of 
$\Gamma_N$ ($N=3,4,5,7,8,16$). As one can
check in the case of $N=3,5$ and $N=7$ all three-dimensional representations of $SL(2,Z_N)$ are also representations of $\Gamma_N$, while
for $N=2,4,8$ only half of the triplets of $SL(2,Z_N)$ also satisfies the relations in eq.(\ref{modular}). Thus, we consider only 19 instead
of 33 triplets.
We list an explicit realization of the generators $S$ and $T$ for each $\Gamma_N$ in this subsection and argue why one such set is sufficient for deducing $S$ and $T$ for all irreducible faithful 
three-dimensional representations. In doing so, we choose a basis in which
the generator $T$ is represented by a diagonal matrix, see \cite{eholzer}.

In the case of $\Gamma_3 \simeq A_4$, we choose the representation matrices $\rho(S)$ and $\rho(T)$ for its irreducible triplet as 
\be
\rho(S)=\frac{1}{3}
\left(
\begin{array}{ccc}
-1&2&2\\
2&-1&2\\
2&2&-1
\end{array}
\right) \,\,\, , \,\,
\rho(T)=\left(
\begin{array}{ccc}
e^{2 \pi i/3}&0&0\\
0&e^{4 \pi i/3}&0\\
0&0&1
\end{array}
\right) 
\ee
which satisfy the defining relations of $A_4$
\be
S^2 = E \, , \,\, (ST)^3 = E  \, , \,\,  T^3 = E \, .
\ee

The group $\Gamma_4 \simeq S_4$ contains two  inequivalent irreducible triplet representations and we choose  $\rho(S)$ and $\rho(T)$ for one of them to be
\be
\label{STS4}
\rho(S)=\frac{1}{2}
\left(
\begin{array}{ccc}
0&\sqrt{2}&\sqrt{2}\\
\sqrt{2}&-1&1\\
\sqrt{2}&1&-1
\end{array}
\right) \,\,\, , \,\,
\rho(T)=\left(
\begin{array}{ccc}
1&0&0\\
0&e^{\pi i/2}&0\\
0&0&e^{3 \pi i/2}
\end{array}
\right) 
\ee
fulfilling the defining relations of $S_4$
\be
\label{relS4}
S^2 = E \, , \,\, (ST)^3 = E \, , \,\,  T^4 = E  \, .
\ee
The generators of the second triplet are related to those in eq.(\ref{STS4}) by a sign change, because the relations in eq.(\ref{relS4}) are invariant under
$\{S, T\} \rightarrow \{-S,-T\}$. It is noteworthy that the group $S_4$ is isomorphic to $\Delta (24)$, i.e. it belongs to the series of $SU(3)$ subgroups 
$\Delta(6 n^2)$ with $n=2$. In appendix \ref{appB} we explicitly show how the generators $S$ and $T$ 
are related to the generators $a$, $b$, $c$ and $d$ used in \cite{D96,Bovier:1980gc} to define the groups $\Delta(6 n^2)$.

Like $\Gamma_4 \simeq S_4$ also the group $\Gamma_5 \simeq A_5$ contains two inequivalent irreducible triplets
and we choose $\rho(S)$ and $\rho(T)$ for one of them to be
\be
\label{STA5}
\rho(S)=\frac{1}{\sqrt{5}}
\left(
\begin{array}{ccc}
1&\sqrt{2}&\sqrt{2}\\
\sqrt{2}&-\phi&1/\phi\\
\sqrt{2}&1/\phi&-\phi\\
\end{array}
\right) \,\,\, , \,
\rho(T)=\left(
\begin{array}{ccc}
1&0&0\\
0&e^{2 \pi i/5}&0\\
0&0&e^{8 \pi i/5}
\end{array}
\right) 
\ee
with $\phi=(1+\sqrt{5})/2$. As one can check they fulfill the defining relations of $A_5$
\be
\label{relA5}
S^2 = E \, , \,\, (ST)^3 = E \, , \,\, T^5 = E  \, .
\ee
Note that if the set $\{S, T\}$ satisfies the relations in eq.(\ref{relA5}) then also $\{T^2ST^3ST^2, T^2 \}$ does, leading  in the case of a three-dimensional representation 
to a second independent representation. Consequently, a set of representation matrices $\rho(S)$ and $\rho(T)$ for the other triplet
can be immediately deduced from the matrices in eq.(\ref{STA5}). 

$\Gamma_7 \simeq PSL(2,Z_7)$ has two irreducible triplets which are complex conjugated. We can choose for one of them $\rho(S)$ and $\rho(T)$ as
\be
\label{STPSL27}
\rho(S)=\frac{2}{\sqrt{7}}
\left(
\begin{array}{ccc}
s_1&s_2&s_3\\
s_2&-s_3&s_1\\
s_3&s_1&-s_2
\end{array}
\right)
\,\,\, , \,
\rho(T)=\left(
\begin{array}{ccc}
e^{4 \pi i/7}&0&0\\
0&e^{2 \pi i/7}&0\\
0&0&e^{8 \pi i/7}
\end{array}
\right)
\ee
with $s_k=\sin k \pi/7$,
fulfilling the defining relations of $PSL(2,Z_7)$
\be
\label{relPSL27}
S^2 = E \, , \,\, (ST)^3 = E  \, , \,\,  T^7 = E \, , \,\, (ST^{-1}ST)^4= E \, .
\ee
Note again that for this group, as for all groups with $N > 5$, at least one additional relation is necessary in order to render the group finite.

In the cases of $\Gamma_8$ and $\Gamma_{16}$ the admissible (four and eight) triplets are not faithful and thus we actually perform our analysis
using the subgroups of $\Gamma_8$ and $\Gamma_{16}$ which are generated through the triplets. For $\Gamma_8$ the generated subgroup
is of order 96 and can be identified as the group $\Delta(96)$ \cite{D96,Bovier:1980gc}, see appendix  \ref{appB}. The generator relations for $S$ and $T$ are
\be
\label{relD96}
S^2 = E \, , \,\, (ST)^3 =  E \, , \,\,   T^8 = E  \, , \,\, (ST^{-1}ST)^3= E 
\ee
and we choose for one of the four triplets $\rho(S)$ and $\rho(T)$ as
\be
\label{STD96}
\rho(S)=\frac{1}{2}
\left(
\begin{array}{ccc}
0&\sqrt{2}&\sqrt{2}\\
\sqrt{2}&-1&1\\
\sqrt{2}&1&-1
\end{array}
\right)
\,\,\, , \,
\rho(T)=\left(
\begin{array}{ccc}
e^{6\pi i/4}&0&0\\
0&e^{7\pi i/4}&0\\
0&0&e^{3\pi i/4}
\end{array}
\right)
\, .
\ee
Note that for each set $\{S, T\}$ which satisfies the relations in eq.(\ref{relD96}) also $\{-S,-T\}$ and the complex conjugate of $S$ and $T$ are solutions.
This leads to four inequivalent faithful irreducible three-dimensional representations which form two complex conjugated pairs. Apart from these, $\Delta(96)$ has two unfaithful irreducible triplets
which are not of interest in our analysis.

The relevant subgroup of $\Gamma_{16}$ with respect to which its eight triplets are faithful representations has 384 elements and it can be shown to be isomorphic to the 
group $\Delta(384)$. The latter can be defined in terms of two generators $S$ and $T$ which satisfy
\be
\label{relD384}
S^2 = E \, , \,\,  (ST)^3 = E \, , \,\,  T^{16} = E  \, , \,\, (ST^{-1}ST)^3= E \, .
\ee
We choose $\rho(S)$ and $\rho(T)$ for one of the triplets to be
\be
\label{STD384}
\rho(S)=\frac{1}{2}
\left(
\begin{array}{ccc}
0&\sqrt{2}&\sqrt{2}\\
\sqrt{2}&-1&1\\
\sqrt{2}&1&-1
\end{array}
\right)
\,\,\, , \,
\rho(T)=\left(
\begin{array}{ccc}
e^{14\pi i/8}&0&0\\
0&e^{5\pi i/8}&0\\
0&0&e^{13\pi i/8}
\end{array}
\right)
\, .
\ee
For each set $\{S, T\}$ which satisfies the relations in eq.(\ref{relD384}) also the sets $\{S, T^n\}$ with $n$ odd (and smaller than 16) 
and $\{ -S, -T\}$ fulfill these relations.\footnote{Note that also the set $\{ -S, -T^n\}$, $n$ odd, fulfills the relations in eq.(\ref{relD384}).}
As consequence, we find eight
such triplet representations and also an explicit set of representation matrices $\rho(g_i)$. The triplets can be grouped into four complex conjugate pairs. 
Apart from these the group also contains six unfaithful irreducible triplets which we do not use in the present study.

In the following section we discuss in detail the lepton mixing originating if the three generations of left-handed lepton doublets are assigned to one of the triplet representations introduced 
in this subsection.

\section{Results for lepton mixing}
\label{lepton_mixing}

We present our results for the mixing patterns which arise, if one of the groups discussed in the preceding section plays the role
of the flavour symmetry $G_l$. We apply the following constraints in our classification: 
\begin{itemize}
\item[a)] left-handed leptons transform as faithful 
irreducible triplet $\rho$ of the group $G_l$, 
\item[b)]  neutrinos are Majorana particles,
\item[c)] the group $G_l$ is broken to $G_e$ and to $G_\nu$ in the charged lepton and neutrino sectors,
respectively. As a consequence the charged lepton and the neutrino mass matrices are invariant under the action of the elements $g_{e i}$ of $G_e$ and $g_{\nu i}$ of
$G_\nu$\footnote{The charged lepton mass matrix $m_l$ is given in the right-left basis.}
\be
\rho(g_{e i})^\dagger m^\dagger_l m_l \rho(g_{e i})=m^\dagger_l m_l~~~,~~~~~~~~~~\rho(g_{\nu i})^T m_\nu~ \rho(g_{\nu i})=m_\nu~~~,
\label{AB}
\ee
\item[d)] $G_\nu$ is constrained to be (contained in) a Klein group by condition b). We choose the transformation properties of $\rho$ under
$Z_2\times Z_2$ such that it decomposes into three inequivalent singlets. This allows to distinguish the three generations. 
A degeneracy in the charges which are assigned to the three generations of lepton doublets would prevent us 
from determining the mixing pattern only through $G_e$, $G_\nu$ and $G_l$.
\item[e)] $G_e$ is taken to be a cyclic group $Z_M$ with index $M\geq 3$ or, if necessary,
\footnote{We constrain ourselves always to the smallest symmetry required to achieve three distinct charged lepton masses; for example, if a $Z_4$ subgroup of $G_l$ can do the
job, we do not consider cases in which $Z_4$ is replaced by a product of cyclic groups containing this $Z_4$ group. A consequence of this procedure might be that we exclude
cases because they do not fulfill condition f), although they can be made fulfilling the latter through an extension of the subgroup considered.} 
a product of cyclic groups, e.g. $Z_2 \times Z_2$. We discard non-abelian residual symmetries $G_e$, since their non-abelian 
character would result in a complete or partial degeneracy of the mass spectrum. For the same reason as in the case of $G_\nu$, we require that $\rho$
decomposes into three inequivalent representations under $G_e$.
\item[f)] we only discuss cases in which the elements of $G_\nu$ and $G_e$ give rise to
the original group $G_l$ and not to one of its subgroups.
\end{itemize}

Lepton mixing originates then from the mismatch of the embedding of $G_e$ and $G_\nu$ into $G_l$ as can be seen in the following way:
we can diagonalize the matrices $\rho(g_{e i})$ and $\rho(g_{\nu i})$ with two unitary transformations $\Omega_e$ and $\Omega_\nu$
\be
\label{diagGeGnu}
\rho(g_{e i})_{diag}=\Omega_e^\dagger~ \rho(g_{e i})~ \Omega_e~~~,~~~~~~~\rho(g_{\nu i})_{diag}=\Omega_\nu^\dagger~ \rho(g_{\nu i})~ \Omega_\nu~~~,
\ee
because both, $G_e$ and $G_\nu$, are abelian. The matrices $\Omega_e$ and $\Omega_\nu$ are determined uniquely up to diagonal unitary matrices $K_{e,\nu}$ and
permutation matrices $P_{e,\nu}$, respectively,
\be
\Omega_{e} \;\; \rightarrow  \;\; \Omega_e~P_e~K_e \;\;\;\;\;\; \mbox{and} \;\;\;\;\;\;  \Omega_{\nu} \;\; \rightarrow  \;\; \Omega_\nu~P_\nu~K_\nu \, .
\ee
From the requirement in eq.(\ref{AB}) it follows that $\Omega_e$ also diagonalizes $m_l^\dagger m_l$ and $\Omega_\nu$ the neutrino mass matrix $m_\nu$.
Thus, the lepton mixing matrix $U_{PMNS}$ is,  up to Majorana phases and permutations of rows and columns, 
\be
\label{UPMNS}
U_{PMNS}=\Omega_e^\dagger\Omega_\nu \, .
\ee
The mixing matrix $U_{PMNS}$ is thus determined through $G_e$ and $G_\nu$ and their relative embedding into $G_l$. However, it is determined only up to exchanges of rows and columns,
because we do not predict lepton masses in this approach.  Hence, the mixing angles are fixed up to a small number of degeneracies, associated with
these possible exchanges. Also the Dirac CP phase $\delta_{CP}$ is determined up to $\pi$, if the exchange of rows and columns is taken into account. At the same time, Majorana
phases cannot be predicted, because they are related to the eigenvalues of the matrix $m_\nu$ which remain unconstrained in this framework.

Switching the roles of $G_e$ and $G_\nu$ obviously leads to $U_{PMNS}$ being hermitian conjugated, see eq.(\ref{UPMNS}).
Note that if a pair of groups $G'_e$ and $G'_\nu$ is conjugated to the pair of groups $G_e$ and $G_\nu$ under the element
$g$ belonging to $G_l$, both pairs lead to the same result for $U_{PMNS}$.
Indeed, if $\Omega_e$ and $\Omega_\nu$ diagonalize the elements of $G_e$ and $G_\nu$, respectively,
 $\rho(g) \Omega_e$ and $\rho(g) \Omega_\nu$ diagonalize 
those of $G'_e$ and $G'_\nu$.
In the preceding section an explicit realization for one of the faithful irreducible three-dimensional representations of each relevant group is given and it is shown how
a realization can be obtained for the other triplets of the group as well. The relevant transformations are the multiplication of the generators with a sign, their complex conjugation
or  taking suitable products of the generators of the given triplet in the case of $A_5$ and $\Delta(384)$. Since the matrices $\rho(g_{ei})$ and $\rho(g_{\nu i})$ appear
twice in the relations shown in eq.(\ref{AB}) it is obvious that a multiplication with a sign does not change the result for $U_{PMNS}$. Regarding the application of complex conjugation
we observe that eq.(\ref{diagGeGnu}) is valid for $\rho(g_{e (\nu) i})^\star$ as well, if we replace $\Omega_{e (\nu)}$ with $\Omega_{e (\nu)}^\star$ and
$\rho(g_{e (\nu) i})_{diag}$ with $\rho(g_{e (\nu) i})_{diag}^\star$. As result also $U_{PMNS}$ has to be complex conjugated, see eq.(\ref{UPMNS}). This does not change the
mixing angles, but $\delta_{CP}$ by $\pi$, see eqs.(\ref{JCP}) and (\ref{JCP2}). In the case of the non-trivial relations between the triplets of $A_5$ and
$\Delta(384)$ one can show that these lead to the same set of representation matrices $\{\rho(g_{i})\}$ (see appendix \ref{appB}). Thus, from the fact that we perform a comprehensive
study of all possible lepton mixing patterns for one particular triplet follows that we find all the possible mixing patterns which can be derived in our framework independently of the
choice of the faithful irreducible three-dimensional representation.

In the following we discuss the different candidates for $G_l$. Since we are mainly interested in the mixing angles we only present
the absolute values of the matrix entries, for which we introduce the notation $||U_{PMNS}||$. We use the freedom in exchanging rows and columns of
$U_{PMNS}$ related to the ordering of the lepton masses in order to present the configuration for which the mixing angles are as close as possible to their 
experimental best fit values; especially, we choose the smallest entry to be the element $U_{e3}$. Furthermore, we choose $|U_{e1}|$ to be larger than or equal
to $|U_{e2}|$ such that the solar mixing angle is smaller or equal to maximal mixing. Similarly, we select the ordering of the second and third rows in the presentation of
the matrix $||U_{PMNS}||$ in such a way that
the resulting atmospheric mixing angle satisfies $\sin^2 \theta_{23} \leq 1/2$ because its best fit value quoted in the global fit \cite{fogli} is below 1/2.\footnote{Two issues
should be noted: first, other global fits \cite{schwetz} foresee a best fit value of $\sin^2\theta_{23}$ slightly larger than 1/2 and, second, the errors of $\sin^2\theta_{23}$ are non-Gaussian
and tolerate better values larger than the best fit value, thus values larger than 1/2.}  However, in the text we also mention the result for $\sin^2 \theta_{23}$ for $|U_{\mu 3}|> |U_{\tau 3}|$.
The Jarlskog invariant $J_{CP}$ is calculated 
as \cite{jcp}
\be
\label{JCP}
J_{CP}=\mathrm{Im} (V_{11} \, V_{12} ^{\star} \, V_{21} ^{\star} \, V_{22})
 = \mathrm{Im} (V_{11} \, V_{13} ^{\star} \, V_{31} ^{\star} \, V_{33})
 = \mathrm{Im} (V_{22} \, V_{23} ^{\star} \, V_{32} ^{\star} \, V_{33}) 
\ee
for a mixing matrix $V$ and can be written in terms of the mixing angles $\theta_{ij}$ and the Dirac CP phase $\delta_{CP}$ as
\be
\label{JCP2}
J_{CP}= \frac{1}{8} \, \sin 2\theta_{12} \, \sin 2\theta_{23} \, \sin 2\theta_{13} \, \cos \theta_{13} \, \sin \delta_{CP} \, .
\ee

\mathversion{bold}
\subsection{Mixing patterns from $A_4$}
\mathversion{normal}

The group $A_4$ has 12 elements which are distributed into four conjugacy classes: $1 \, {\cal C}_1$, $3 \, {\cal C}_2$, $4 \, {\cal C}^1_3$ and $4 \, {\cal C}^2_3$
with $a \, {\cal C}_b$ denoting a class with $a$ (distinct) elements which have order $b$. Note the first class $1 \, {\cal C}_1$ is always the
trivial one which only contains the neutral element $E$ of the group. Having four conjugacy classes, $A_4$ possesses four irreducible representations: three singlets,
the trivial one and a complex conjugated pair, and the triplet. 
The subgroups of $A_4$ are $Z_2$, $Z_3$ and $Z_2 \times Z_2$ and only the latter two are relevant for us. 
A representative of the classes $a \, {\cal C}_b$,  written in terms of the generators $S$ and $T$ is:
\be
1 \, {\cal C}_1 : \, E \,\, , \, 3 \, {\cal C}_2 : \,  S \,\, , \, 4 \, {\cal C}^1_3  : \, T \,\, , \, 4 \, {\cal C}^2_3 : \, T^2 \, .
\ee
The Klein group $K$ is generated by the elements $S$, $T^2ST$ and the four different $Z_3$ subgroups $C_i$ by $T$, $ST$, $TS$ and $STS$, respectively.
The latter are all conjugate. We have a unique choice for the group $G_\nu$ and four different ones for $G_e$. Obviously, in all cases the elements of
$G_\nu$ and $G_e$ generate $A_4$.

We find a unique mixing pattern
\be
\label{eq:magic}
||U_{PMNS}||=
\frac{1}{\sqrt{3}}
\left(
\begin{array}{ccc}
1 & 1 & 1 \\
1 & 1 & 1 \\
1 & 1 &1
\end{array}
\right)
\ee
which predicts both solar and atmospheric mixing angles to be maximal and $\theta_{13}$ to fulfill $\sin^2 \theta_{13}=1/3$.
This pattern also leads to a maximal Dirac CP phase $|\delta_{CP}|=\pi/2$ and thus $|J_{CP}|=1/(6\sqrt{3})\approx 0.096$. 
Obviously, $\theta_{12}$ and $\theta_{13}$ need large corrections in order to be compatible with experimental data.
These results have also been found in \cite{Lam11}. The mixing pattern in eq.(\ref{eq:magic}) has been discussed often in the literature, see \cite{magic78}.

The reason why we do not find tri-bimaximal mixing for $G_e=Z_3$ and $G_\nu=Z_2 \times Z_2$, but instead the pattern in eq.(\ref{eq:magic}), is that in order to achieve
the former pattern the symmetry preserved in the neutrino sector is composed of a $Z_2$ contained in $A_4$ and an accidental $Z_2$ group. This can easily happen in explicit models, if a certain
choice of flavour symmetry breaking fields is made \cite{A4examples}.

\mathversion{bold}
\subsection{Mixing patterns from $S_4$}
\mathversion{normal}

The group $S_4$ has 24 elements and five conjugacy classes: $1 \, {\cal C}_1$,  $3 \, {\cal C}_2$, $6 \, {\cal C}_2$, $8 \, {\cal C}_3$ and $6 \, {\cal C}_4$.
The five irreducible representations corresponding to these classes are two singlets, one doublet and two triplets.
Its abelian subgroups are $Z_2$, $Z_3$, $Z_4$ and $Z_2 \times Z_2$. 
A representative for each of the classes can be written in terms of the generators $S$ and $T$:
\be\nonumber
1 \, {\cal C}_1 : \, E \,\, , \, 3 \, {\cal C}_2 : \,  T^2 \,\, , \, 6 \, {\cal C}_2 : \, S \,\, , \, 8 \, {\cal C}_3 : \, ST \,\, , \, 6 \, {\cal C}_4 : \, T \, .
\ee
We find four different Klein groups: $K$ which can be generated through $T^2$ and $ST^2S$, $K_1$ given
by $S$, $T^2ST^2$, $K_2$ by $T^2$, $ST^2ST$ and $K_3$ by $ST^2S$, $T^3ST$. As one can check $K$ is a normal subgroup of $S_4$ and all $K_i$ are conjugate to each other. We find
four distinct $Z_3$ subgroups $C_i$ which are generated by the following elements: $ST$, $TS$, $T^2ST$ and $TST^2$, respectively. Note all of them are conjugate. The three
different $Z_4$ subgroups $Q_i$ are generated by: $T$, $T^2S$ and $STS$, respectively. Like the $Z_3$ subgroups, the $Z_4$ groups are all conjugate to each other.

When computing the mixing patterns we can distinguish three possibilities for $G_e$, namely $G_e=Z_3$, $G_e=Z_4$ or $G_e=Z_2\times Z_2$. In the latter case $G_e$ and
$G_\nu$ obviously cannot be the same Klein group. We discuss the different cases in turn. Note that in this occasion our requirement to generate the whole group $S_4$ through the 
generators of $G_e$ and $G_\nu$ excludes some of the possible combinations, especially if the normal Klein group $K$ is involved. 

\mathversion{bold}
\subsubsection*{$G_\nu=Z_2\times Z_2$ ~~~~~ $G_e=Z_3$}
\mathversion{normal}

Using these subgroups we are able to derive only tri-bimaximal mixing
\be
\label{eq:TBM}
||U_{PMNS}||=\frac{1}{\sqrt{6}}
\left(
\begin{array}{ccc}
2&\sqrt{2}&0\\
1&\sqrt{2}&\sqrt{3}\\
1&\sqrt{2}&\sqrt{3}\\
\end{array}
\right)~~~,
\ee
if we apply the constraint that the generators of $G_\nu$ and $G_e$ should give rise to $S_4$.  An example of $G_\nu$ and $G_e$ is $G_\nu=K_1$ and $G_e=C_3$.
The mixing parameters are then $\sin^2\theta_{23}=1/2$, $\sin^2\theta_{12}=1/3$, vanishing $\theta_{13}$ and $J_{CP}=0$. According to recent experimental indications \cite{T2K,MINOS,DC}  very small 
values of $\theta_{13}$ are disfavoured and thus models in which tri-bimaximal mixing is realized should contain sources of corrections to the prediction $\theta_{13}=0$ so that
$\theta_{13} \sim 0.1 \div 0.2$ can be accommodated.

\mathversion{bold}
\subsubsection*{$G_\nu=Z_2\times Z_2$ ~~~~~ $G_e=Z_4$}
\mathversion{normal}

Also in this case we find a unique mixing pattern, this time bimaximal mixing
\be
\label{eq:bimax}
||U_{PMNS}||=\frac{1}{2}
\left(
\begin{array}{ccc}
\sqrt{2}&\sqrt{2}&0\\
1&1&\sqrt{2}\\
1&1&\sqrt{2}\\
\end{array}
\right)~~~,
\ee
applying the constraint that the generators of $G_\nu$ and $G_e$ should give rise to $S_4$.  An example of $G_\nu$ and $G_e$ is $G_\nu=K_1$ and $G_e=Q_3$.
The mixing angles are then $\sin^2\theta_{23}=\sin^2\theta_{12}=1/2$, vanishing $\theta_{13}$. We find again $J_{CP}=0$. In this case, the solar as well as the reactor mixing 
angle have to undergo appropriate corrections in order to be in accordance with experimental data. Such corrections have to be of the order $0.1\div0.2$, which
is roughly the size of the Cabibbo angle, see e.g. \cite{S4BM}.

\mathversion{bold}
\subsubsection*{$G_\nu=Z_2\times Z_2$ ~~~~~ $G_e=Z_2\times Z_2$}
\mathversion{normal}

Also in this case only one pattern can be produced, restricting ourselves to cases in which the whole group $S_4$ is generated, which is again bimaximal
mixing as shown in eq.(\ref{eq:bimax}). One possible choice of $G_e$ and $G_\nu$ is $G_e=K_1$ and $G_\nu=K_2$.

\vspace{0.1cm}

\noindent Note that all our results coincide with those of \cite{Lam11}.

\mathversion{bold}
\subsection{Mixing patterns from $A_5$}
\mathversion{normal}

The group $A_5$ has 60 elements and five conjugacy classes: $1 \, {\cal C}_1$,  $15 \, {\cal C}_2$, $20 \, {\cal C}_3$, $12 \, {\cal C}_5^1$ and $12 \, {\cal C}_5^2$.
The latter indicate the existence of five irreducible representations: the trivial singlet, two triplets, one four- and one five-dimensional representation.
The abelian subgroups are $Z_2$, $Z_3$, $Z_5$ and $Z_2 \times Z_2$. 
Also here we can give a representative of each class in terms of $S$ and $T$:
\be\nonumber
1 \, {\cal C}_1 : \, E \,\, , \, 15 \, {\cal C}_2 : \, S  \,\, , \, 20 \, {\cal C}_3 : \,  ST \,\, , \, 12 \, {\cal C}_5^1 : \, T \,\, , \, 12 \, {\cal C}_5^2 : \,  T^2 \,  .
\ee
The generating elements of the five distinct Klein groups $K_i$,  the six $Z_5$ subgroups $R_i$ and the ten different $Z_3$ subgroups $C_i$ can be found in table \ref{tab_abel_A5} in appendix \ref{appA}. 
Note that all of them are conjugate to each other.
Three possible cases can be discussed: $G_e$ is a $Z_3$, $Z_5$ or a Klein group, while $G_\nu=Z_2\times Z_2$ is fixed.

\mathversion{bold}
\subsubsection*{$G_\nu=Z_2\times Z_2$ ~~~~~ $G_e=Z_3$}
\mathversion{normal}

We get the unique mixing pattern
\be
\label{eq:A5mix1}
||U_{PMNS}||=
\frac{1}{\sqrt{6}}
\left(
\begin{array}{ccc}
 \sqrt{2} \phi  & \sqrt{2}/\phi  & 0 \\
 1/\phi  & \phi  & \sqrt{3} \\
 1/\phi  & \phi  & \sqrt{3}
\end{array}
\right)
\approx
\left(
\begin{array}{ccc}
 0.934 & 0.357 & 0 \\
 0.252 & 0.661 & 0.707 \\
 0.252 & 0.661 & 0.707 \\
\end{array}
\right)~~~,
\ee
if we require condition f) to be fulfilled.
This mixing pattern is generated for $G_\nu=K_1$ and $G_e=C_1$. The mixing angles are vanishing $\theta_{13}$ and maximal $\theta_{23}$ together with $\sin^2\theta_{12}=\frac{1}{3}(2-\phi)=
\frac{1}{6}(3-\sqrt{5})\approx 0.127$. Obviously, $J_{CP}=0$. In this case,  especially $\theta_{12}$ has to acquire large corrections in order to match the experimental
best fit value. This pattern is also found in \cite{Lam11}.

\mathversion{bold}
\subsubsection*{$G_\nu=Z_2\times Z_2$ ~~~~~ $G_e=Z_5$}
\mathversion{normal}

In this case all combinations of $G_\nu$ and $G_e$ lead to the mixing pattern
\be
\label{eq:A5mix2}
||U_{PMNS}||=
\left(
\begin{array}{ccc}
\cos \theta_{12} & \sin \theta_{12} & 0 \\
 \sin \theta_{12} /\sqrt{2} & \cos \theta_{12} /\sqrt{2} & 1/\sqrt{2} \\
 \sin \theta_{12} /\sqrt{2} & \cos \theta_{12} /\sqrt{2} & 1/\sqrt{2}
\end{array}
\right)
\approx
\left(
\begin{array}{ccc}
 0.851 & 0.526 & 0 \\
 0.372 & 0.602 & 0.707 \\
 0.372 & 0.602 & 0.707
\end{array}
\right)~~
\ee
with $\tan\theta_{12}=1/\phi$. Again, we find vanishing $\theta_{13}$ and maximal atmospheric mixing $\theta_{23}$
together with $\sin^2 \theta_{12}\approx 0.276$ and $J_{CP}=0$. Moderate corrections to $\theta_{12}$ and $\theta_{13}$ are necessary in order to achieve 
agreement with the experimental data. Such a pattern has been discussed in \cite{GR}.

\mathversion{bold}
\subsubsection*{$G_\nu=Z_2\times Z_2$ ~~~~~ $G_e=Z_2\times Z_2$}
\mathversion{normal}

If both sectors are invariant under a Klein group, the unique mixing matrix is
\be
\label{eq:A5mix3}
||U_{PMNS}||=
\frac{1}{2}
\left(
\begin{array}{ccc}
\phi & 1 &1/ \phi\\
1/\phi & \phi & 1 \\
1 &1/\phi& \phi
\end{array}
\right)
\approx
\left(
\begin{array}{ccc}
 0.809 & 0.5 & 0.309 \\
  0.309 & 0.809 & 0.5 \\
   0.5 & 0.309 & 0.809
\end{array}
\right)~~~.
\ee
Excluding the case in which $G_\nu$ and $G_e$ are the same group, we always find the pattern in eq.(\ref{eq:A5mix3}) to be generated.
The mixing angles extracted from eq.(\ref{eq:A5mix3}) are: $\sin^2\theta_{13}\approx 0.095$ and $\sin^2 \theta_{12} =\sin^2 \theta_{23}=\frac{1}{10} (5-\sqrt{5})\approx 0.276$.
Also in this case there is no non-trivial Dirac CP phase, i.e. $J_{CP}=0$. By exchanging the second and third rows in the pattern in eq.(\ref{eq:A5mix3})
we can find another reasonable pattern predicting $\sin^2\theta_{23}=\frac{1}{10}(5+\sqrt{5}) \approx 0.724$. In this case, none of the three mixing angles
is in the experimentally preferred $2 \sigma$ range \cite{fogli} and thus in a model, realizing this pattern at LO, all have to receive considerable corrections to make the model viable. 

\mathversion{bold}
\subsection{Mixing patterns from $PSL(2,Z_7)$}
\mathversion{normal}

The group $PSL(2,Z_7)$ has 168 elements and six conjugacy classes: $1 \, {\cal C}_1$, $21 \, {\cal C}_2$,  $56 \, {\cal C}_3$, $42 \, {\cal C}_4$, $24 \, {\cal C}_7^1$
and $24 \, {\cal C}_7^2$. According to its six classes, the group also has six irreducible representations: 
one singlet, one six-, one seven-, one eight-dimensional representation as well as a pair of complex conjugated triplets. 
The abelian subgroups relevant for us are $Z_3$, $Z_4$, $Z_2 \times Z_2$ and $Z_7$. 
We give one representative for each class:
\be\nonumber
1 \, {\cal C}_1  : \, E \,\, , \,   21 \, {\cal C}_2  : \,  S  \,\, , \,  56 \, {\cal C}_3 : \,  ST  \,\, , \,  42 \, {\cal C}_4  : \, ST^3  \,\, , \,  24 \, {\cal C}_7^1  : \, T  \,\, , \,  24 \, {\cal C}_7^2 :\, T^3 \, .
\ee
The elements of order 2 form 14 Klein groups $K_i$, while the other elements generate 28 $Z_3$, 21 $Z_4$ and eight $Z_7$ symmetries, which we call $C_i$, $Q_i$ and $P_i$ in the following, respectively. 
The cyclic subgroups are conjugate to each other, while the Klein groups can be divided into two categories. A list of elements generating them can be found in table \ref{tab_abel_PSL27} in appendix \ref{appA}.
We find four possible cases to discuss: $G_e$ is a $Z_3$, $Z_4$, $Z_7$ or a Klein group, while $G_\nu$ is fixed to be a Klein group through the Majorana nature of the three
neutrinos.

\mathversion{bold}
\subsubsection*{$G_\nu=Z_2\times Z_2$ ~~~~~ $G_e=Z_3$}
\mathversion{normal}

If the neutrino sector is invariant under a Klein group and the charged lepton sector under an element of order 3, we find as unique mixing pattern
\bea
\nonumber
||U_{PMNS}||&=&\frac{1}{\sqrt{6}}
\left(
\begin{array}{ccc}
 \sqrt{\frac{1}{2} \left( 5 + \sqrt{21}\right)} & 1 & \sqrt{\frac{1}{2} (5-\sqrt{21})}\\
 1 & 2 & 1\\
 \sqrt{\frac{1}{2} (5-\sqrt{21})} & 1 &  \sqrt{\frac{1}{2} \left( 5 + \sqrt{21}\right)}
\end{array}
\right)
\\ \label{eq:mixP1}
&\approx&
\left(
\begin{array}{ccc}
 0.894 & 0.408 & 0.187 \\
  0.408 & 0.816 & 0.408\\
 0.187 & 0.408 & 0.894 
\end{array}
\right)~~~
\eea
which satisfies all requirements a)-f). One possible choice of $G_\nu$ and $G_e$ is: $G_\nu=K_1$ and $G_e=C_1$.
The mixing angles are: $\sin^2 \theta_{13}=\frac{1}{12} \left( 5 -\sqrt{21}\right) \approx 0.035$ 
and $\sin^2 \theta_{12}=\sin^2 \theta_{23}=\frac{1}{14} \left( 7-\sqrt{21}\right)\approx 0.173$. 
The CP violating phase fulfills $|\sin \delta_{CP}|= \sqrt{7/8}\approx 0.935$ and thus $|J_{CP}|=1/(24 \sqrt{3})\approx 0.024$.
Exchanging the second and third rows
in eq.(\ref{eq:mixP1}) we get $\sin^2\theta_{23}=\frac{1}{14}(7+\sqrt{21})\approx 0.827$. In both cases however, the atmospheric as well as
the solar mixing angle necessitate large corrections in order to be compatible with experimental data and the only feature of the LO which reflects the
data well is the small value of $\theta_{13}$.

\mathversion{bold}
\subsubsection*{$G_\nu=Z_2\times Z_2$ ~~~~~ $G_e=Z_4$}
\mathversion{normal}

Requiring the conditions a)-f) to be fulfilled, the only admissible pattern is
\bea
\nonumber
||U_{PMNS}|| &=&\frac{1}{2}
\left(
\begin{array}{ccc}
 \sqrt{\frac{1}{2} \left( 3+\sqrt{7}\right)} &  1 & \sqrt{\frac{1}{2} \left( 3-\sqrt{7}\right)} \\
 1 & \sqrt{2} & 1\\
 \sqrt{\frac{1}{2} \left( 3-\sqrt{7}\right)} & 1 &  \sqrt{\frac{1}{2} \left( 3+\sqrt{7}\right)}
\end{array}
\right)
\\ \label{eq:mixP2}
&\approx&\left(
\begin{array}{ccc}
 0.840 &  0.5 & 0.210 \\
 0.5 & 0.707 & 0.5 \\
  0.210 & 0.5 & 0.840
\end{array}
\right)~~~.
\eea
It is produced, for example, for $G_\nu=K_1$ and $G_e=Q_1$.
We extract as mixing angles: $\sin^2 \theta_{13}=\frac{1}{8} \left( 3-\sqrt{7}\right)\approx 0.044$ and  $\sin^2 \theta_{23}=\sin^2 \theta_{12}=\frac{1}{9} \left( 5-\sqrt{7}\right)\approx 0.262$.
The Jarlskog invariant fulfills $|J_{CP}|=1/32\approx 0.031$ and hence $|\sin\delta_{CP}|=\frac{1}{4}\sqrt{13-\sqrt{7}}\approx 0.804$.
If the second and third rows of the matrix in eq.(\ref{eq:mixP2}) are exchanged, 
we find $\sin^2\theta_{23}=\frac{1}{9}(4+\sqrt{7})\approx 0.738$. Again, the atmospheric mixing angle requires large corrections, in order to be in the experimentally preferred range, 
while moderate ones are sufficient for the solar mixing angle.

\mathversion{bold}
\subsubsection*{$G_\nu=Z_2\times Z_2$ ~~~~~ $G_e=Z_7$}
\mathversion{normal}

In this case the mixing matrix takes the form
\be
 \label{eq:mixP3}
||U_{PMNS}||=2 \sqrt{\frac{2}{7}}
\left(
\begin{array}{ccc}
 s_2 s_3 & s_1 s_3 & s_1 s_2 \\
 s_1 s_2 & s_2 s_3 & s_1 s_3 \\
 s_1 s_3 & s_1 s_2 & s_2 s_3
\end{array}
\right)
\approx
\left(
\begin{array}{ccc}
 0.815 & 0.452 & 0.363 \\
 0.363 & 0.815 & 0.452 \\
  0.452 & 0.363 & 0.815
\end{array}
\right)~~~
\ee
and the mixing parameters read: $\sin^2 \theta_{13}\approx 0.132$, $\sin^2 \theta_{12}=\sin^2 \theta_{23} \approx 0.235$ and $|J_{CP}|=1/(8\sqrt{7})\approx 0.047$.
This pattern arises from any possible combination of a Klein group and an element of order seven. Exchanging the second and third rows we achieve
$\sin^2\theta_{23}\approx 0.765$. Since the reactor mixing angle is not particularly small and $\theta_{23}$ needs large corrections, this pattern does not seem
to be suited as LO one in a model in which corrections are mild.

\mathversion{bold}
\subsubsection*{$G_\nu=Z_2\times Z_2$ ~~~~~ $G_e=Z_2\times Z_2$}
\mathversion{normal}

If both, the neutrino and the charged lepton, sectors are invariant under a Klein group, the unique mixing pattern, compatible with our requirements, is
\be
\label{eq:mixP4}
||U_{PMNS}||=
\frac{1}{2}
\left(
\begin{array}{ccc}
\sqrt{2} & 1 & 1 \\
 1 &\sqrt{2} & 1 \\
 1 & 1 & \sqrt{2}
\end{array}
\right)~~~.
\ee
We find for the mixing angles and $J_{CP}$: $\sin^2\theta_{12}=\sin^2\theta_{23}=1/3$, $\sin^2\theta_{13}=1/4$ and $|J_{CP}|=\sqrt{7}/32\approx 0.083$.
The quantity $|\sin\delta_{CP}|$ is thus $3\sqrt{7}/8 \approx 0.992$.  
One choice of $G_\nu$ and $G_e$ leading to this particular pattern is $G_\nu=K_1$ and $G_e=K_3$. The other reasonable value for $\theta_{23}$
arises, if second and third rows are exchanged in eq.(\ref{eq:mixP4}), namely $\sin^2\theta_{23}=2/3$. Also this pattern does not predict a small value of $\theta_{13}$ 
and $\theta_{23}$ is compatible with the data only at $3\sigma$ level \cite{fogli}.

\mathversion{bold}
\subsection{Mixing patterns from $\Delta(96)$}
\mathversion{normal}

The group $\Delta(96)$ has 96 elements and ten conjugacy classes: $1 \, {\cal C}_1$, $3 \, {\cal C}_2$, 
$12 \, {\cal C}_2$, $32 \, {\cal C}_3$, $3 \, {\cal C}_4^1$, $3 \, {\cal C}_4^2$, $6 \, {\cal C}_4$, $12 \, {\cal C}_4$, $12 \, {\cal C}_8^1$ and $12 \, {\cal C}_8^2$.
Thus, it has also ten irreducible representations which are two singlets, one doublet, six triplets and one six-dimensional representation.
The abelian subgroups relevant for our purposes are $Z_3$, $Z_4$, $Z_2 \times Z_2$, $Z_8$ as well as the products $Z_2 \times Z_4$ and $Z_4 \times Z_4$ (see below).
We give one representative for each of the classes $a \, {\cal C}_b$ in terms of $S$ and $T$:
\bea\nonumber
&& 1 \, {\cal C}_1  : \, E \,\, , \,   3 \, {\cal C}_2   : \,  T^4  \,\, , \,   12 \, {\cal C}_2     : \,  S  \,\, , \,  32 \, {\cal C}_3   : \,  ST  \,\, , \,    3 \, {\cal C}_4^1   : \,   T^2 \,\, , \,   
 3 \, {\cal C}_4^2   : \,   T^6  \,\, , \,   
\\ \nonumber  
&&  6 \, {\cal C}_4     : \,  ST^2ST^4  \,\, ,
 \,  12 \, {\cal C}_4   : \,  ST^4  \,\, , \,  12 \, {\cal C}_8^1    : \,  T  \,\, , \,     12 \, {\cal C}_8^2   : \,  T^3    \, .
\eea
A list of generating elements for the abelian subgroups $Z_3$,  $Z_4$, $Z_2 \times Z_2$ and $Z_8$ can be found in table \ref{tab_abel_D96} and that of the 15 $Z_2$
symmetries $V_i$ in table \ref{tab_Z2_D96} in appendix \ref{appA}.
Note that the six Klein groups $K_i$ are conjugate to each other, while $K$ is a normal subgroup of $\Delta(96)$. 
The 16 $Z_3$ subgroups $C_i$ are conjugate to each other. The same holds for all six $Z_8$ subgroups $O_i$. 
The twelve $Z_4$ subgroups $Q_i$ fall into three categories applying similarity transformations belonging to $\Delta(96)$: the first contains $Q_1$, $Q_2$ and $Q_3$,
the second one $Q_4$, $Q_5$ and $Q_6$ and the third the others $Q_7$, ..., $Q_{12}$. 

In the following we discuss all possible combinations of $G_e$ and $G_\nu=Z_2 \times Z_2$ case by case. We encounter two new instances: first, we come across situations in which 
a certain combination of types of subgroups employed for $G_e$ and $G_\nu=Z_2 \times Z_2$ does not allow to generate the original group $\Delta(96)$, see e.g. $G_e=Z_4$ and $G_\nu=Z_2 \times Z_2$,
and second, it can be checked that the generating elements of the groups $Q_1$, $Q_2$ and $Q_3$  for the faithful irreducible triplets (to which we assign the 
left-handed lepton doublets) are represented by matrices which have two degenerate eigenvalues. As a consequence, it is not possible to distinguish among the three 
generations of leptons with these groups and the latter cannot be used  as $G_e$. However, we can use them, if we consider $G_e$ to be $Z_2 \times Z_4$ or $Z_4 \times Z_4$.
In the first case, $Z_4$ is one of the three groups $Q_{1,2,3}$ and $Z_2$ is one of the 15 distinct $Z_2$ symmetries $V_i$ contained in $\Delta(96)$.
In the second case, $G_e=Z_4 \times Z_4$, both $Z_4$ groups are generated through elements associated with matrices having two degenerate eigenvalues for the irreducible faithful triplets. However,
this degeneracy is resolved, if the product $Z_4 \times Z_4$ is considered. Indeed,  all three products $Q_1 \times Q_2$, $Q_1 \times Q_3$ and $Q_2 \times Q_3$ allow to resolve this degeneracy
and are admissible because they give rise to a group of order 16. For this reason we also include the cases $G_e=Z_2 \times Z_4$ and $G_e=Z_4 \times Z_4$ in our discussion.\footnote{We do not discuss all possible types of $Z_2 \times Z_4$ and $Z_4 \times Z_4$ subgroups, but only those in which the $Z_4$ group alone is not sufficient for distinguishing the
three generations of leptons. This procedure is in agreement with our requirement e).}

\mathversion{bold}
\subsubsection*{$G_\nu=Z_2\times Z_2$ ~~~~~ $G_e=Z_3$}
\mathversion{normal}

We find as admissible mixing pattern, if the neutrino sector is invariant under a Klein group and the charged lepton sector under an element of order 3,
\be
\label{eq:mixD96}
||U_{PMNS}||=
\frac{1}{\sqrt{3}}
\left(
\begin{array}{ccc}
\frac{1}{2}(\sqrt{3}+1) & 1 & \frac{1}{2}(\sqrt{3}-1) \\
   1                    & 1 & 1 \\
\frac{1}{2}(\sqrt{3}-1) & 1 & \frac{1}{2}(\sqrt{3}+1)
\end{array}
\right)~~~
\approx
\left(
\begin{array}{ccc}
 0.789 &  0.577 &0.211\\
 0.577  & 0.577 & 0.577 \\
 0.211 &  0.577 &0.789
\end{array}
\right)
\ee
leading to the following mixing angles: $\sin^2\theta_{23}=\sin^2\theta_{12}=\frac{8-2\sqrt{3}}{13}\approx 0.349$ and $\sin^2\theta_{13}=\frac{2-\sqrt{3}}{6}\approx 0.045$ 
together with $J_{CP}=0$. A viable choice of $G_e$ and $G_\nu$ leading to this pattern
is $G_e=C_1$ and $G_\nu=K_1$. This pattern has already been discussed in \cite{dATFH11}, because it is one of the few patterns originating from a non-trivial breaking of a flavour symmetry 
which leads to  small $\theta_{13}$, as indicated by the latest experimental data \cite{T2K,MINOS,DC}. At the same time, the solar and atmospheric mixing angles are compatible with
the preferred values from global fits \cite{fogli} at the $3\sigma$ level. A second possibility for the value of $\sin^2\theta_{23}$ is $\sin^2\theta_{23}=\frac{1}{13}(5+2\sqrt{3})\approx 0.651$, as already mentioned in 
\cite{dATFH11}. According to \cite{dATFH11} we call the mixing pattern arising from the matrix in eq.(\ref{eq:mixD96}) {\tt M2} and the one originating from the exchange of second and
third rows in eq.(\ref{eq:mixD96})  {\tt M1}.

\mathversion{bold}
\subsubsection*{$G_\nu=Z_2\times Z_2$ ~~~~~ $G_e=Z_4$}
\mathversion{normal}

In this case none of the choices of generating elements of $G_e$ and $G_\nu$ allows us to recover the group $\Delta(96)$ itself, but only one of its proper subgroups.

\mathversion{bold}
\subsubsection*{$G_\nu=Z_2\times Z_2$ ~~~~~ $G_e=Z_8$}
\mathversion{normal}

If the original group $\Delta(96)$ is generated through the elements of $G_e$ and $G_\nu$, the resulting mixing pattern predicts bimaximal mixing, see eq.(\ref{eq:bimax}).
One possible choice of $G_e$ and $G_\nu$ is $G_e=O_1$ and $G_\nu=K_1$.

\mathversion{bold}
\subsubsection*{$G_\nu=Z_2\times Z_2$ ~~~~~ $G_e=Z_2\times Z_2$}
\mathversion{normal}

Similar to the case in which $G_e$ is a $Z_4$ subgroup we cannot generate the original group $\Delta(96)$ with the elements of the subgroups $G_e$ and $G_\nu$ in this case.

\mathversion{bold}
\subsubsection*{$G_\nu=Z_2\times Z_2$ ~~~~~ $G_e=Z_2 \times Z_4$}
\mathversion{normal}

As discussed above, we consider the $Z_4$ subgroup contained in $G_e$ to be one of the subgroups whose elements are represented by matrices with degenerate eigenvalues
for the irreducible faithful triplet.
Then only for bimaximal mixing, see eq.(\ref{eq:bimax}), all our requirements are passed. An example of $G_e$ and $G_\nu$ is:
$G_e=V_1 \times Q_1$ and $G_\nu=K_1$.

\mathversion{bold}
\subsubsection*{$G_\nu=Z_2\times Z_2$ ~~~~~ $G_e=Z_4 \times Z_4$}
\mathversion{normal}

Assuming that only $Z_4$ subgroups $Q_{1,2,3}$ are allowed we see that for none of the possible choices we can generate the original group $\Delta(96)$
using the elements of the subgroups $G_e$ and $G_\nu$.

\mathversion{bold}
\subsection{Mixing patterns from $\Delta(384)$}
\mathversion{normal}

The group $\Delta(384)$ has 384 elements and 24 conjugacy classes: $1 \, {\cal C}_1$, $3 \, {\cal C}_2$, $24 \, {\cal C}_2$, 
$128 \, {\cal C}_3$, $3 \, {\cal C}_4^1$, $3 \, {\cal C}_4^2$, $6 \, {\cal C}_4$, $24 \, {\cal C}_4$, $3 \, {\cal C}^i_8$, $6 \, {\cal C}^j_8$, $24 \, {\cal C}^k_8$, $24 \, {\cal C}^i_{16}$ with
$i=1,...,4$, $j=1,...,6$ and $k=1,2$. Its 24 representations are:  two singlets, one doublet, 14 triplets as well as seven six-dimensional representations.
The abelian subgroups relevant for our purposes are $Z_3$, $Z_4$, $Z_2 \times Z_2$, $Z_8$, $Z_{16}$ as well as the products $Z_2 \times Z_4$, $Z_2\times Z_8$, $Z_4 \times Z_4$, 
$Z_4 \times Z_8$ and $Z_8 \times Z_8$ (see below).
We give a representative for each of the classes $a \, {\cal C}_b$:
\bea\nonumber
&& 1 \, {\cal C}_1    : \, E \,\, , \,    3 \, {\cal C}_2    : \,   T^8   \,\, , \,        24 \, {\cal C}_2    : \,  S    \,\, , \,      128 \, {\cal C}_3    : \,   ST  \,\, , \,       3 \, {\cal C}_4^1    : \,   T^4  \,\, , \,       3 \, {\cal C}_4^2     : \,  T^{12}   \,\, , \,   
6 \, {\cal C}_4    : \,  ST^8ST^4   \,\, , \,  
\\ \nonumber
&&    24 \, {\cal C}_4    : \,  ST^8  \,\, , \,      3 \, {\cal C}^1_8   : \,  T^2   \,\, , \,     3 \, {\cal C}^2_8      : \,  T^6   \,\, , \,      3 \, {\cal C}^3_8      : \,  T^{10}   \,\, , \,         3 \, {\cal C}^4_8    : \,    T^{14} \,\, , \,   6 \, {\cal C}^1_8    : \,  ST^2ST^{10}   \,\, , \, 
\\ \nonumber
&&    6 \, {\cal C}^2_8     : \,  ST^2ST^6   \,\, , \,       6 \, {\cal C}^3_8     : \,   ST^2ST^{12}   \,\, , \,        6 \, {\cal C}^4_8      : \,  ST^2ST^4    \,\, , \,       6 \, {\cal C}^5_8      : \,  ST^4ST^{10}    \,\, , \,        6 \, {\cal C}^6_8    : \,  ST^2ST^8   \,\, , \,   
\\ \nonumber
&& 24 \, {\cal C}^1_8    : \,  ST^4    \,\, , \,        24 \, {\cal C}^2_8     : \,  ST^{12}   \,\, , \,     24 \, {\cal C}^1_{16}     : \,  T   \,\, , \,      24 \, {\cal C}^2_{16}      : \,  T^3   \,\, , \,      24 \, {\cal C}^3_{16}        : \,  T^5   \,\, , \,      24 \, {\cal C}^4_{16}  : \,  T^7   \, .  
\eea
A list of the generating elements of $Z_2\times Z_2$, $Z_3$,  $Z_4$, $Z_8$, $Z_{16}$ and $Z_2$ can be found in the tables \ref{tab_Klein_D384}-\ref{tab_Z2_D384} listed in appendix \ref{appA}. 
The twelve Klein groups $K_i$ are conjugate to each other, while $K$ is a normal subgroup of $\Delta(384)$. 
The 64 $Z_3$ groups $C_i$ and the 12 $Z_{16}$ groups $Y_i$
are conjugate to each other, respectively. Furthermore, the 18 $Z_4$ groups $Q_i$ fall into three categories whose members are conjugate to each other: $Q_{1,2,3}$ form one category, $Q_{4,5,6}$ a second 
one and the remaining 12 groups $Q_i$, $i=7,...,18$ the third category. Similarly, the 24 $Z_8$ groups $O_i$ can be divided into five such categories: the first one containing $O_{1,2,3}$, the second
one $O_{4,5,6}$, the third one $O_{7,8,9}$, the forth one $O_{10,11,12}$ and the last one the remaining 12 groups $O_i$ with $i=13,...,24$. 
The 27 $Z_2$ groups are denoted by $V_i$.

Similar to what happens in the case of the group $\Delta(96)$ also here the matrices $\rho(g)$ representing the $Z_4$ generating elements of the groups $Q_1$, $Q_2$, $Q_3$
and the ones representing the $Z_8$ generating elements of the groups $O_1$, $O_2$, $O_3$ have two degenerate eigenvalues such that none of these groups can play the 
role of $G_e$. However, as explained in the preceding subsection products containing these groups can be used as $G_e$. We consider the following cases in our discussion:
$G_e=Z_2 \times Z_4$, $G_e=Z_2 \times Z_8$,  $G_e=Z_4  \times Z_4$,  $G_e=Z_4 \times Z_8$ and  $G_e=Z_8 \times Z_8$ with each of the two factors alone being insufficient to distinguish among the three
generations. The latter requirement is imposed in order to meet condition e). We checked that in all cases considered the distinction among the three generations becomes possible
and that the group is indeed of the correct order $m \cdot n$, as expected for a product $Z_m \times Z_n$.

\mathversion{bold}
\subsubsection*{$G_\nu=Z_2\times Z_2$ ~~~~~ $G_e=Z_3$}
\mathversion{normal}

The mixing pattern, compatible with the requirements stated above, is
\bea\nonumber
||U_{PMNS}||&=&
\frac{1}{\sqrt{3}}
\left(
\begin{array}{ccc}
\frac{1}{2}\sqrt{4+\sqrt{2}+\sqrt{6}}&1& \frac{1}{2}\sqrt{4-\sqrt{2}-\sqrt{6}}\\
\frac{1}{2}\sqrt{4+\sqrt{2}-\sqrt{6}}&1&\frac{1}{2}\sqrt{4-\sqrt{2}+\sqrt{6}}\\
\sqrt{1-\frac{1}{\sqrt{2}}}&1&\sqrt{1+\frac{1}{\sqrt{2}}}
\end{array}
\right)\\
\label{eq:mixD384}
&\approx&
\left(
\begin{array}{ccc}
0.810 & 0.577 & 0.107 \\
0.497 & 0.577 & 0.648 \\
0.312 & 0.577 & 0.754
\end{array}
\right)~~~.
\eea
It arises, for example, if we choose $G_e=C_1$ and $G_\nu=K_1$. The values of the mixing angles are:
$\sin^2\theta_{12}=\frac{4}{8+\sqrt{2}+\sqrt{6}}\approx 0.337$, 
$\sin^2\theta_{23}=\frac{4-\sqrt{2}+\sqrt{6}}{8+\sqrt{2}+\sqrt{6}}\approx 0.424$ and
$\sin^2\theta_{13}=\frac{4-\sqrt{2}-\sqrt{6}}{12}\approx 0.011$. The Jarlskog invariant $J_{CP}$ vanishes. Switching the second
and third rows of the matrix in eq.(\ref{eq:mixD384}) leads to 
$\sin^2\theta_{23}=\frac{4+2\sqrt{2}}{8+\sqrt{2}+\sqrt{6}}\approx 0.576$.
This has already been reported in \cite{dATFH11} and the mixing associated with the matrix in eq.(\ref{eq:mixD384}) has been denoted by pattern {\tt M3}, while the
pattern {\tt M4} is realized, if second and third rows are exchanged in eq.(\ref{eq:mixD384}).
Both patterns, {\tt M3} and {\tt M4}, accommodate the experimental results very well and it depends on the used global fit \cite{fogli,maltoni} or \cite{schwetz} which of the 
two patterns is preferred over the other.

\mathversion{bold}
\subsubsection*{$G_\nu=Z_2\times Z_2$ ~~~~~ $G_e=Z_4$}
\mathversion{normal}

As one can check, in all cases $G_\nu=K$ or $G_\nu=K_i$ and $G_e=Q_j$ the elements of $G_\nu$ and $G_e$ do not give rise to the entire group
$\Delta(384)$, but only one of its proper subgroups is generated.

\mathversion{bold}
\subsubsection*{$G_\nu=Z_2\times Z_2$ ~~~~~ $G_e=Z_8$}
\mathversion{normal}

Also in this case the requirement to generate the whole group through the elements of the subgroups $G_\nu$ and $G_e$ cannot be fulfilled for any choice of Klein group for $G_\nu$ and $Z_8$ group for  $G_e$.

\mathversion{bold}
\subsubsection*{$G_\nu=Z_2\times Z_2$ ~~~~~ $G_e=Z_{16}$}
\mathversion{normal}

Requiring the conditions a)-f) to be fulfilled leads to one admissible category of combinations of $G_\nu$ and $G_e$ which  gives rise to bimaximal mixing, see eq.(\ref{eq:bimax}).
For example, we can take $G_\nu=K_1$ and $G_e=Y_1$.

\mathversion{bold}
\subsubsection*{$G_\nu=Z_2\times Z_2$ ~~~~~ $G_e=Z_2\times Z_2$}
\mathversion{normal}

In the case of both groups, $G_\nu$ and $G_e$, being a Klein group we cannot reproduce the original group $\Delta(384)$ and thus we do not consider such combinations as admissible.

\mathversion{bold}
\subsubsection*{$G_\nu=Z_2\times Z_2$ ~~~~~ $G_e=Z_2 \times Z_4$}
\mathversion{normal}

Imposing requirement f) we see that this case is not admissible.

\mathversion{bold}
\subsubsection*{$G_\nu=Z_2\times Z_2$ ~~~~~ $G_e=Z_2 \times Z_8$}
\mathversion{normal}

We find one admissible category of choices of $G_\nu$ and $G_e$. The associated mixing pattern is the bimaximal one, see eq.(\ref{eq:bimax}). One possible choice
of $G_\nu$ and $G_e$ is: $G_\nu=K_1$ and $G_e=V_1 \times O_1$.

\mathversion{bold}
\subsubsection*{$G_\nu=Z_2\times Z_2$ ~~~~~ $G_e=Z_4 \times Z_4$}
\mathversion{normal}

All choices of $G_\nu=K$ or $G_\nu=K_i$ and $G_e$ being equal to one of the three products, $Q_1 \times Q_2$, $Q_1 \times Q_3$ and $Q_2\times Q_3$, do not meet requirement f) and thus
are not considered.

\mathversion{bold}
\subsubsection*{$G_\nu=Z_2\times Z_2$ ~~~~~ $G_e=Z_4 \times Z_8$}
\mathversion{normal}

Also in this case all possible choices of $G_\nu$ and $G_e$ in accordance with requirement e) do not fulfill requirement f).

\mathversion{bold}
\subsubsection*{$G_\nu=Z_2\times Z_2$ ~~~~~ $G_e=Z_8 \times Z_8$}
\mathversion{normal}

Finally, an inspection of the possible choices of $G_\nu$ and $G_e$ which fall into the present category shows that none is compatible with all requirements imposed at the beginning of this 
section.

\mathversion{bold}
\section{Selective results for mixing patterns arising from general $G_\nu$}
\mathversion{normal}

In this section we abandon the requirement that neutrinos are Majorana particles and thus $G_\nu$ is no longer fixed to be a Klein group, but can be any abelian
group which allows to distinguish among the three generations as in the case of $G_e$. We do not intend to repeat a comprehensive study as for $G_\nu=Z_2 \times Z_2$
and instead only highlight some interesting results and new mixing patterns which we find.

The simplest extension of our results, which are obtained under the assumption that $G_\nu$ is a Klein group, is to consider the case in which the roles of $G_e$ and $G_\nu$ are switched, i.e.
$G_e$ is a Klein group and $G_\nu$ an arbitrary subgroup. A quick inspection of the transposes of the solutions $||U_{PMNS}||$ presented in the preceding section however shows that 
they do not give rise to any particularly interesting result for the mixing angles.

We are able to produce the mixing pattern in eq.(\ref{eq:mixD96}) associated with the group $\Delta(96)$, if we consider $S_4$ as flavour group and
choose $G_e=Z_3$ and $G_\nu=Z_4$. This result does not depend on a particular choice of the $Z_3$ and $Z_4$ groups. This observation is interesting because it allows us
to produce a pattern with small $\theta_{13}$ with the help of the well-known group $S_4$ instead of using $\Delta(96)$.

For $G_l=A_5$ we find that it is possible to generate a pattern in which the solar as well as the atmospheric mixing angle are
maximal and at the same time the reactor mixing angle is in the experimentally preferred range. Explicitly, we get
\bea\nonumber
||U_{PMNS}||
&=&
\left(
\begin{array}{ccc}
\sqrt{\frac{1}{15} (5+\sqrt{5})}&\sqrt{\frac{1}{15} (5+\sqrt{5})}&\frac{3-\sqrt{5}}{\sqrt{6(5-\sqrt{5}})}\\
\frac{1}{2}\left(1-\sqrt{\frac{1}{15} (5-2\sqrt{5})} \right) &\frac{1}{2}\left(1+\sqrt{\frac{1}{15} (5-2\sqrt{5})} \right)&\sqrt{\frac{1}{15} (5+\sqrt{5})}\\
\frac{1}{2}\left(1+\sqrt{\frac{1}{15} (5-2\sqrt{5})}\right)&\frac{1}{2}\left(1-\sqrt{\frac{1}{15} (5-2\sqrt{5})} \right)&\sqrt{\frac{1}{15} (5+\sqrt{5})}
\end{array}
\right)
\\ 
\label{eq:BMth13}
&\approx&
\left(
\begin{array}{ccc}
0.695 & 0.695 & 0.188 \\
0.406 & 0.594 & 0.695\\
0.594 & 0.406 & 0.695
\end{array}
\right) \, .
\eea
The reactor mixing angle is $\sin^2 \theta_{13}= \frac{\sqrt{5}-2}{3\sqrt{5}}\approx 0.0352$ and $J_{CP}$ vanishes. 
This pattern requires that we choose $G_e=Z_3$ and $G_\nu=Z_5$ and furthermore, that we make a specific choice, e.g. an admissible one is $G_e=C_{10}$ and $G_\nu=R_1$.

Also in the case of the group $PSL(2,Z_7)$ another potentially interesting pattern can be found, if the condition that neutrinos are Majorana particles is abandoned.
Taking $G_e$ to be a $Z_4$ subgroup and $G_\nu$ to be a $Z_7$ group we get
\be
\label{eq:mixPadd}
||U_{PMNS}||
\approx
\left(
\begin{array}{ccc}
0.850 & 0.519 & 0.090 \\
0.382 & 0.725 & 0.573\\
0.363 & 0.452 & 0.815
\end{array}
\right) \, .
\ee
The mixing angles which can be extracted are: $\sin^2\theta_{23} \approx 0.331$, $\sin^2\theta_{12} \approx 0.272$
and $\sin^2\theta_{13} \approx 0.0081$. Again, $J_{CP}=0$. Alternatively, we can exchange the second and third rows in the matrix in eq.(\ref{eq:mixPadd}) so that
$\sin^2\theta_{23}\approx 0.669$.

Similar to the fact that we are able to generate the mixing pattern associated with the group $\Delta(96)$ also with the group $S_4$, if we assume neutrinos to be Dirac particles instead of 
Majorana ones, we are able to generate the pattern in eq.(\ref{eq:mixD384}) associated with the group $\Delta(384)$ also with the group $\Delta(96)$. This can be achieved for $G_e=Z_3$
and $G_\nu=Z_8$. Note that the pattern does not depend on the particular choice of the groups $Z_3$ and $Z_8$ and that the elements of
the subgroups $G_e$ and $G_\nu$ always give rise to the original group $\Delta(96)$.

Considering the group $\Delta(384)$ and $G_e$ to be a $Z_3$ subgroup, while $G_\nu$ is a $Z_{16}$ subgroup, we find two interesting mixing patterns. First, 
\bea\nonumber
||U_{PMNS}||
&=& \frac{1}{\sqrt{6}}
\left(
\begin{array}{ccc}
 \sqrt{2+\sqrt{2+\sqrt{2}}} & \sqrt{2} & \sqrt{2-\sqrt{2 +\sqrt{2}}}\\
\sqrt{2-\sqrt{2-\sqrt{2+\sqrt{3}}}} & \sqrt{2} & \sqrt{2+\sqrt{2-\sqrt{2+\sqrt{3}}}}\\
\sqrt{2-\sqrt{2+\sqrt{2-\sqrt{3}}}} & \sqrt{2} & \sqrt{2+\sqrt{2+\sqrt{2-\sqrt{3}}}}
\end{array}
\right)
\\  \label{eq:D384gen1}
&\approx&
\left(
\begin{array}{ccc}
0.801 & 0.577 & 0.159 \\
0.538 & 0.577 & 0.614\\
0.262 & 0.577 & 0.773
\end{array}
\right) 
\eea
which leads to $\sin^2 \theta_{12}= \frac{2}{4+\sqrt{2+\sqrt{2}}} \approx 0.342$, 
$\sin^2 \theta_{23}= \frac{2+\sqrt{2-\sqrt{2+\sqrt{3}}}}{4+\sqrt{2+\sqrt{2}}} \approx 0.387$ 
and $\sin^2 \theta_{13}= \frac{1}{6}(2-\sqrt{2+\sqrt{2}}) \approx 0.025$ as well as $J_{CP}=0$. We denote this pattern by {\tt M5}.
The other admissible value of $\sin^2\theta_{23}$ is $\sin^2\theta_{23}= \frac{2+\sqrt{2+\sqrt{2-\sqrt{3}}}}{4+\sqrt{2+\sqrt{2}}} \approx 0.613$ and the
associated pattern is called {\tt M6}.
The second pattern reads
\bea\nonumber
||U_{PMNS}||
&=& \frac{1}{\sqrt{6}}
\left(
\begin{array}{ccc}
 \sqrt{2+\sqrt{2+\sqrt{2+\sqrt{3}}}} & \sqrt{2} & \sqrt{2-\sqrt{2 +\sqrt{2+\sqrt{3}}}}\\
\sqrt{2-\sqrt{2-\sqrt{2}}} & \sqrt{2} & \sqrt{2+\sqrt{2-\sqrt{2}}}\\
\sqrt{2-\sqrt{2-\sqrt{2-\sqrt{3}}}} & \sqrt{2} & \sqrt{2+\sqrt{2-\sqrt{2-\sqrt{3}}}}
\end{array}
\right)
\\  \label{eq:D384gen2}
&\approx&
\left(
\begin{array}{ccc}
0.815 & 0.577 & 0.053 \\
0.454 & 0.577 & 0.679\\
0.361 & 0.577 & 0.732
\end{array}
\right) 
\eea
from which follows $\sin^2 \theta_{12}= \frac{2}{4+\sqrt{2+\sqrt{2+\sqrt{3}}}} \approx 0.334$, 
$\sin^2 \theta_{23}= \frac{2+\sqrt{2-\sqrt{2}}}{4+\sqrt{2+\sqrt{2+\sqrt{3}}}} \approx 0.462$ 
and $\sin^2 \theta_{13}= \frac{1}{6}(2-\sqrt{2+\sqrt{2+\sqrt{3}}}) \approx 0.0029$ as well as $J_{CP}=0$.
We call this pattern {\tt M7}.
Again, exchanging the second and third rows in eq.(\ref{eq:D384gen2}) gives rise to another value of 
$\sin^2\theta_{23}=\frac{4+\sqrt{2(4+\sqrt{2}-\sqrt{6})}}{8+2\sqrt{2+\sqrt{2+\sqrt{3}}}} \approx 0.538$ and the pattern is denoted by {\tt M8}.
The first pattern can be generated with $G_e$ chosen as $C_1$ and $G_\nu$ as $Y_5$, while
the second pattern arises, for example, from the choice $G_e=C_1$ and $G_\nu=Y_1$. Note that only these two patterns can be
generated if we choose $G_e$ to be a $Z_3$ group and $G_\nu$ a $Z_{16}$ subgroup.

\begin{figure}[t!]
\begin{center}
\includegraphics[height=12cm]{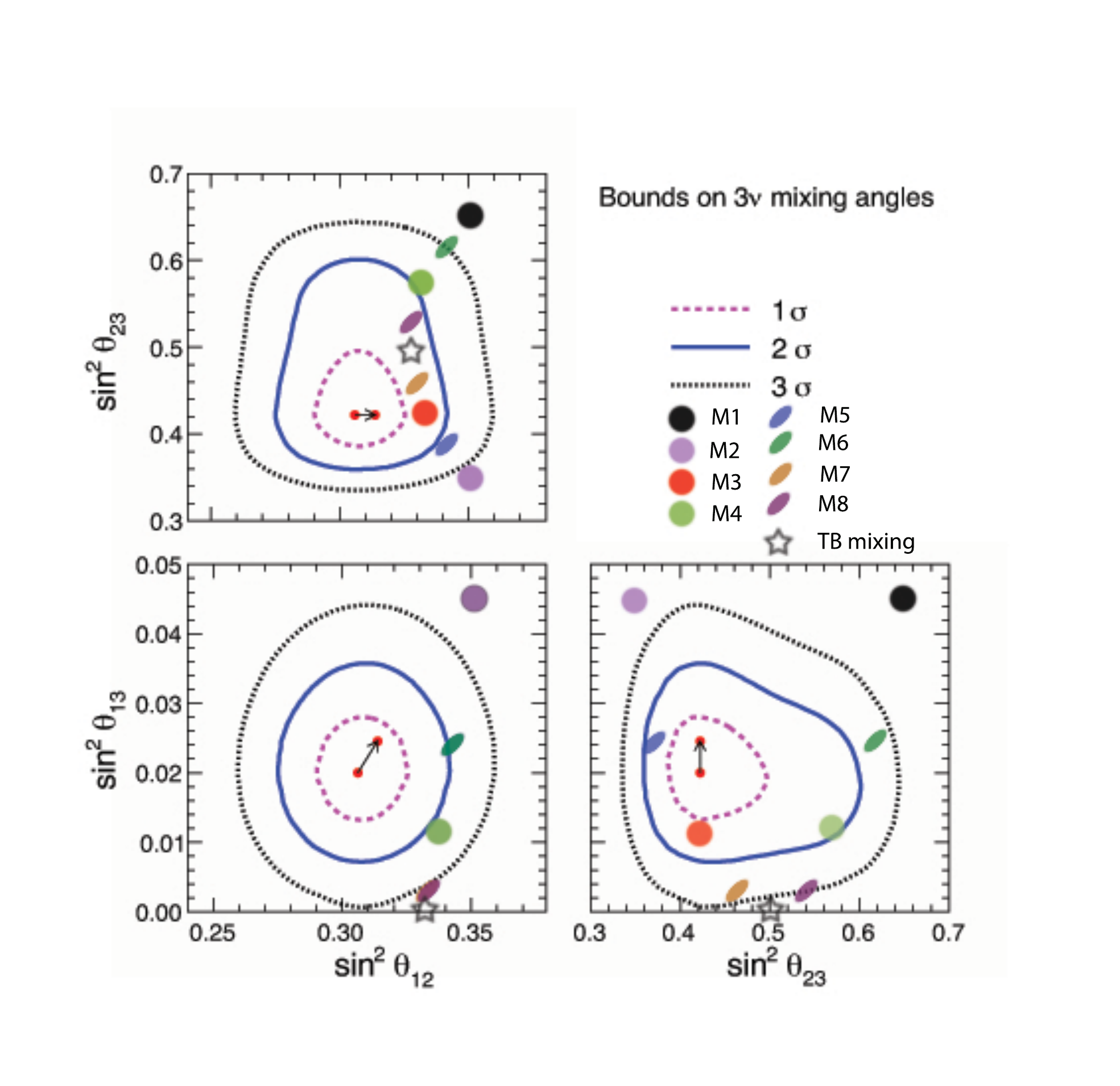} 
\caption{Values of $\sin^2 \theta_{ij}$ for the mixing patterns {\tt M1}, ..., {\tt M8} and for tri-bimaximal mixing (marked with an open star)
which can all be written in the form of eq.(\ref{eq:TBgen}) for different values of $\alpha$. The patterns {\tt M1}, ..., {\tt M4}
(represented by dots) arise for neutrinos being Majorana or Dirac particles, while the patterns {\tt M5}, ..., {\tt M8} (represented by ellipses) 
require Dirac neutrinos. 
The counters show the
$1 \sigma$ (pink dashed line), $2 \sigma$ (blue solid line) and $3 \sigma$ (black dotted line) levels and are taken from 
\cite{fogli}. The small dots indicate the best fit values of the mixing angles and the arrows the effect of the new estimates
of the reactor antineutrino flux. 
Note that in the $\sin^2 \theta_{12}$-$\sin^2 \theta_{13}$ plane the points of \texttt{M1} and \texttt{M2}, of {\tt M3} and {\tt M4}, of {\tt M5} and {\tt M6} as well
as of \texttt{M7} and \texttt{M8} lie on top of each other, since they only differ in the value of $\sin^2 \theta_{23}$. \label{figure_alpha}}
\end{center}
\end{figure}
As has already been shown in \cite{dATFH11}, the patterns in eqs. (\ref{eq:mixD96}) and (\ref{eq:mixD384}) can be parametrized as particular deviations from tri-bimaximal mixing,\footnote{For a discussion
of such deviations from tri-bimaximal mixing with $\alpha$ arbitrary see \cite{alphaarb}.} 
 i.e.
\be
\label{eq:TBgen}
U_{PMNS}=\left( \begin{array}{ccc}
 \sqrt{\frac{2}{3}} & \frac{1}{\sqrt{3}} & 0\\
 -\frac{1}{\sqrt{6}} & \frac{1}{\sqrt{3}} & \frac{1}{\sqrt{2}}\\
 -\frac{1}{\sqrt{6}} & \frac{1}{\sqrt{3}} & -\frac{1}{\sqrt{2}}
\end{array}
\right)
\, U_{13}(\alpha)~~~~\mathrm{with}~~~~
U_{13}(\alpha)=
\left(
\begin{array}{ccc}
\cos\alpha&0&\sin\alpha\\
0&1&0\\
-\sin\alpha&0&\cos\alpha
\end{array}
\right)
\ee
with $\alpha= \pm \pi/12$ for the pattern in eq.(\ref{eq:mixD96}) and $\alpha=\pm \pi/24$ for the one in eq.(\ref{eq:mixD384}). Similarly, one can show that the pattern
in eq.(\ref{eq:D384gen1}) is of this particular form with $\alpha=\pm \pi/16$. The pattern mentioned in eq.(\ref{eq:D384gen2}) is of the form in eq.(\ref{eq:TBgen}),
if $\alpha=\pm \pi/48$. In all cases the choice of the sign of $\alpha$ corresponds to the exchange of the second and third rows. All patterns have
in common that the Dirac CP phase $\delta_{CP}$ is trivial. We present these patterns and their compatibility with experimental data \cite{fogli} graphically in figure
\ref{figure_alpha}.

One might hypothesize four possible general formulae for mixing patterns depending on the index $n$ of the group $\Delta(6 n^2)$
which we have explicitly confirmed for $n=2,4,8$. First we consider the case in which $G_\nu$ is fixed to be a Klein group, because 
neutrinos are assumed to be Majorana particles. We find as possible general formulae
\be
\alpha=\frac{\pi}{n} \;\;\; \mbox{for} \;\;\; G_e =Z_3 \;\; , \; G_\nu=Z_2 \times Z_2 
\ee
and
\be
\alpha=\frac{\pi}{3 n} \;\;\; \mbox{for} \;\;\; G_e =Z_3 \;\; , \; G_\nu=Z_2 \times Z_2 \, .
\ee
Note that for $n=2$ we find the matrix $||U_{PMNS}||$ to be equal to the tri-bimaximal one as regards the absolute value, if rows and columns of the former are appropriately permuted. Note further
that in all three cases $n=2,4,8$ the formulae give rise to the same matrix $||U_{PMNS}||$ up to permutations of rows and columns. In the case of 
arbitrary $G_\nu$, implying that neutrinos are Dirac particles, we can consider the two general formulae
\be
\alpha=\frac{\pi}{2 n} \;\;\; \mbox{for} \;\;\; G_e =Z_3 \;\; , \; G_\nu=Z_{2n}
\ee
and 
\be
\alpha=\frac{\pi}{6 n} \;\;\; \mbox{for} \;\;\; G_e =Z_3 \;\; , \; G_\nu=Z_{2n} \, .
\ee
These only lead in the case $n=8$ to two distinct patterns, while for $n=2$ and $n=4$ the patterns are the same up to exchanges of rows and columns.

\section{Comments on quark mixing}

One might ask whether it is also possible to derive quark mixing in the same way, as presented for leptons, assuming that the flavour
group of the quark sector is $G_q$ and that it is broken to $G_d$ and $G_u$ in the down quark and up quark sectors, respectively.\footnote{We assume in
the following that the groups $G_d$, $G_u$ and $G_q$ fulfill the same requirements as $G_e$, $G_\nu$ and $G_l$, see beginning of section
\ref{lepton_mixing}. Obviously, $G_d$ and $G_u$ are not constrained to be Klein groups, because quarks are not Majorana particles.} 
Indeed, one might think of the following possibilities: 
\begin{itemize}
\item[a)] first we can argue, since the CKM mixing matrix is in a rough approximation the identity matrix, 
that this points to a situation in which $G_d$ and $G_u$ are the same and thus no non-trivial mixing arises at LO. Such an idea is, for example, realized
in certain $A_4$ models, see \cite{A4examples}, and in a $T'$ model \cite{FHLM07}.\footnote{Note that in the case of the $T'$ model the requirement that the charges of the three generations of
quarks are different under the subgroups $G_d$ and $G_u$ is not fulfilled and thus the down quark and up quark mass matrices are not diagonal in the limit of unbroken 
$G_d$ and $G_u$. As consequence, one of the mixing angles is generated in the symmetry limit and depends on the size of the entries of the down and up quark mass matrices.} 
\item[b)] we can be more ambitious and can try to explain the (hierarchical) mixing pattern among
quarks, i.e. at least two non-vanishing mixing angles, through the mismatch of the relative embedding of $G_d$ and $G_u$ into the flavour symmetry $G_q$.
In this case left-handed quarks have to be assigned, similar to left-handed leptons, to an irreducible triplet of the group $G_q$. 
\item[c)] the third possibility is given by the fact
that the CKM mixing matrix contains - in rather good approximation - only one non-zero mixing angle, namely the Cabibbo angle. In this case we immediately see that the three
generations of quarks have to be assigned to a two-dimensional and a one-dimensional representation of $G_q$ in order to produce only one non-vanishing
mixing angle (the other two then have to arise from corrections to this LO setup). This can be understood in the following way:
consider a basis in which the matrices representing the elements of $G_u$ are diagonal. 
Then those representing the elements of $G_d$ have to have block-diagonal form with (12) and (21) being the only off-diagonal elements that are non-zero. Since the elements of $G_d$
and $G_u$ are supposed to generate the original flavour group $G_q$ we see that we have found a basis in which the triplet representation, to which the left-handed quarks 
are assigned, decomposes into a two-dimensional (irreducible) and a one-dimensional representation, because the matrices representing the elements of $G_u$ and $G_d$
and as consequence those of $G_q$ are all in block-diagonal form. 
\end{itemize}
The idea to generate the Cabibbo angle according to possibility c) has been proposed in an analysis of dihedral groups 
\cite{BHL07,BH09}, see also \cite{Lam0708}. However,
in this context the requirement that the preserved subgroups $G_d$ and $G_u$ allow for a distinction among the three generations of quarks is frequently not fulfilled, because
$G_d$ and $G_u$ are $Z_2$ subgroups of $G_q$. Thus, the Cabibbo angle is predicted in terms of group-theoretical quantities, 
while the other two mixing angles are not constrained and  depend on the size of non-zero entries of
the up and down quark mass matrices. In certain explicit models 
such unconstrained entries might be zero or suppressed due to a particular choice of flavour symmetry breaking
fields, see e.g. \cite{BH09}, and thus giving rise to the correct order of magnitude of the smaller two quark mixing angles.

In the case of the groups $\Gamma_N$ it is rather simple to comment on the possibilities for quarks: the groups $A_4$, $A_5$ and $PSL(2,Z_7)$ do not have irreducible two-dimensional
representations. Hence, we only have possibilities a) and b) at our disposal; however, we did not find in our study any mixing pattern with two or three small hierarchical mixing angles. Consequently,
we are forced to assume that in models with $G_q$ being $A_4$, $A_5$ or $PSL(2,Z_7)$ quark mixing vanishes at LO in our approach. On the contrary, the groups
$S_4 \simeq \Delta(24)$, $\Delta(96)$ and $\Delta(384)$ have each one irreducible two-dimensional representation. As can be shown, this representation originates from the $S_3$ factor 
present in the semi-direct product $(Z_n\times Z_n)\rtimes S_3$ to which 
the groups $\Delta(6 n^2)$ are isomorphic, see for details appendix \ref{appB} and \cite{D96}. Thus, we might expect similar results 
in all three cases. Concerning the other two possibilities: obviously, possibility a) can always be realized, while we did not find a mixing pattern which favours possibility b).

We can also think of the following situation: if $G_q \subset G_l$ and we are able to generate
the original group $G_l$ with the elements of the subgroups $G_e$ and $G_\nu$, we can abandon the requirement that the original group should be generated also through the elements
of the groups $G_d$ and $G_u$ alone. Then cases in which an irreducible triplet representation of $G_l$ decomposes into a two-dimensional and a one-dimensional representation
under $G_q$ become interesting, because this is an alternative way to realize possibility c). One explicit example of such a case can be found for $G_l=\Delta(384)$
and $G_q$ being a subgroup of $G_l$ of order 128. Taking $G_u=Y_8$ and $G_d=K_1$ we find for the mixing matrix $V_{CKM}$ arising from the mismatch of the embedding of $G_d$ and $G_u$
into $G_q \subset G_l$
\be
||V_{CKM}||= \left(\begin{array}{ccc}
 \cos \pi/16 & \sin \pi/16 & 0\\
 \sin \pi/16 & \cos \pi/16 & 0\\
0 & 0 & 1
\end{array}
\right)
\approx
\left(\begin{array}{ccc}
 0.981 & 0.195 & 0\\
 0.195 & 0.981 & 0\\
 0 & 0 & 1
\end{array}
\right)
\ee
which approximates rather well the experimental best fit values of the elements $|V_{ud}|=0.97428$, $|V_{us}|=0.2253$, $|V_{cd}|=0.2252$ and $|V_{cs}|=0.97345$
\cite{pdg}.

\section{Conclusions}

We have pursued the idea that a discrete flavour symmetry $G_l \subseteq G_f$ is broken to subgroups $G_\nu$ and $G_e$ in the neutrino and charged lepton sectors, respectively, 
determining the lepton mixing pattern as the mismatch of the relative embedding of $G_e$ and $G_\nu$ into $G_l$, up to a small number of degeneracies. We focussed on the series
of finite modular groups $\Gamma_N$ playing the role of $G_l$ and have shown that the requirement of having three-dimensional irreducible representations reduces the number of independent cases
to six groups $\Gamma_N$ with $N=3,4,5,7,8,16$. We have performed a comprehensive study for $G_e$ arbitrary and $G_\nu=Z_2 \times Z_2$, being the maximal invariance 
group for three generations of Majorana neutrinos. Apart from finding well-known patterns such as tri-bimaximal, bimaximal and the golden ratio mixing we have revealed two interesting
patterns predicting $\theta_{13} \sim 0.1 \div 0.2$ \cite{dATFH11} as preferred by the latest experimental results \cite{T2K,MINOS,DC} and global fits \cite{fogli,schwetz,maltoni}. 
We have also presented several promising patterns in the case of $G_\nu$ arbitrary, as is possible, if neutrinos are Dirac particles. Among them are patterns which lead to bimaximal
mixing and small $\theta_{13}$ at the same time, see eq.(\ref{eq:BMth13}), as well as further two patterns with small $\theta_{13}$ and $\theta_{12}$ and $\theta_{23}$ in the 
experimentally preferred range, see eqs.(\ref{eq:D384gen1}) and (\ref{eq:D384gen2}). 
These two together with three patterns found in the case of
$G_\nu=Z_2 \times Z_2$ can be cast into the form given in eq.(\ref{eq:TBgen}) showing that they are specific deviations from tri-bimaximal mixing parametrized with one angle $\alpha$. Their compatibility with
experimental data can be read off from figure \ref{figure_alpha}. We have also commented on the possibilities to derive viable patterns for quark mixing following the idea of preserving non-trivial
subgroups in the down quark and up quark sectors in general as well as focussing on finite modular groups being the flavour symmetry.

\section*{Acknowledgments}

RdAT would like to thank Patrick Ludl for useful discussions.
The work of RdAT is part of the research program of the Dutch Foundation for Fundamental Research of Matter (FOM). RdAT acknowledges 
the hospitality of the University of Padova, where part of this research was completed. FF and CH have been partly supported by the European Programme "Unification in the LHC Era", contract PITN-GA-2009-237920 (UNILHC). CH would like to thank the Aspen Center for Physics for kind hospitality during the preparation of this work.
\newpage
\appendix

\section{Additional information on the presented groups}
\label{appB}

In the first subsection we show that the transformations found in the case of the groups $A_5$ and $\Delta(384)$, respectively, which
relate the generators of the inequivalent faithful irreducible three-dimensional representations lead to the same set of representation matrices. 
In the subsequent subsections additional information on the groups is presented; especially, we prove the equivalence between the definitions of 
the groups $S_4 \simeq \Delta(24)$, $\Delta(96)$ and $\Delta(384)$ given here and the one found in \cite{D96,Bovier:1980gc}.

\subsection*{Equivalence of different triplets with respect to fermion mixing}

Two of the groups considered in our analysis, $A_5$ and $\Delta(384)$, possess inequivalent faithful irreducible three-dimensional representations $\rho_1$ and $\rho_2$
which are related by 
\be
\rho_1(S')=\rho_2(S)~~~~~~~~~~~~~~\rho_1(T')=\rho_2(T)~~~,
\ee
where 
\bea
S'=T^2ST^3ST^2&, &T'=T^2~~~~~~~~~~~~~~~~~~~~~A_5\nn\\
S'=S~~~~~~~~~~~~&, &T'=T^n~,~~~~n~{\rm odd}~~~~~~\Delta(384)~~~.
\eea
Both $\{ S,T\}$ and $\{S',T'\}$ satisfy the relations of the corresponding group, see eqs.(\ref{relA5}) and (\ref{relD384}), respectively.
We show that $\rho_1$ and $\rho_2$ are associated with the same set of representation matrices.

Since both sets $\{S,T\}$ and $\{S',T'\}$ can be chosen as generators of the group $G$, a set of functions $M_i$ $(i=1, ..., |G|)$ exists of the form
\be
M_i(A,B)=\prod_{k=1}^{N^{(i)}} A^{n_k^{(i)}} B^{m_k^{(i)}}~~~~~~~~~~~~(N^{(i)},n_k^{(i)},m_k^{(i)}~{\mbox{non-negative integers}})
\ee
such that the two sets
\be
\{g_i\}=\{M_i(S,T)\}~~~~~~~~~~~\{g'_i\}=\{M_i(S',T')\}
\ee
represent the group $G$. As a  consequence, they have to be equal, up to a permutation of their elements. Therefore 
the sets of matrices $G_1=\{\rho_1(g_i)\}$ and $G_2=\{\rho_1(g'_i)\}$ also coincide.
Furthermore we have 
\be
\rho_2(g_i)=\rho_1(g'_i) \, ,
\ee
because
\bea
\rho_1(g'_i)&=&\rho_1(M_i(S',T')) =M_i(\rho_1(S'),\rho_1(T')) =M_i(\rho_2(S),\rho_2(T))\nn\\
&=&\rho_2(M_i(S,T))=\rho_2(g_i) \, .
\eea
Thus, the sets $G_1=\{\rho_1(g_i)\}$ and $G_2=\{\rho_2(g_i)\}$, explicit realizations of the group $G$ in terms of matrices for the representations $\rho_1$ and $\rho_2$, respectively,
are the same.

\mathversion{bold}
\subsection*{$S_4$}
\mathversion{normal}

We briefly comment on the relation of the generators $S$ and $T$ used in the present paper to define the group $S_4 \simeq \Delta(24)$
and the definition of the groups $\Delta(6 n^2)$ as given in \cite{D96,Bovier:1980gc} which is based on the fact that  
\be
\Delta(6n^2)~\simeq~(Z_n\times Z_n)\rtimes S_3~~,
\ee
i.e. the groups $\Delta(6 n^2)$ are isomorphic to the semi-direct products of $S_3$ and $Z_n \times Z_n$. Thus these can be defined in terms of four generators $a$, $b$, $c$ and $d$
fulfilling the relations
\be
\label{abcd1}
a^3 ~=~ b^2 ~=~ (ab)^2 ~=~ c^n ~=~ d^n ~=~E, 
\ee
\be
\label{abcd2}
cd ~=~ dc ,
\ee
\be
\label{abcd3}
\begin{array}{cccccc}
a c a^{-1}  &~=~&  c^{-1} d^{-1}, \quad
&a d a^{-1} &~=~&  c,             \\ 
b c b^{-1}  &~=~&  d^{-1},        
&b d b^{-1} &~=~&  c^{-1}.      \\    
\end{array}
\ee
As one can see, $a$ and $b$ generate $S_3$ and $c$ and $d$ a $Z_n$ group each.
For $S_4$, we find that $a$, $b$, $c$ and $d$ can be expressed through $S$ and $T$ as
\be
\begin{array}{ll}
a=TST^2~~,~~~~~~~~~~~~&b=S\\
c=T^2~~,~~~~~~~~~~&d=ST^2S
\end{array}~~~~~.
\ee
Such a set of relations is not unique and can only be defined up to similarity transformations applied to the right-hand side of these relations.

\mathversion{bold}
\subsection*{$\Delta(96)$}
\mathversion{normal}

We note that several of the generating elements of the subgroups of $\Delta(96)$ commute, especially
the generators of $Q_{1,...,6}$ all commute. Furthermore it holds that the elements of $Q_1$ commute with those of the groups $Q_9$ and $Q_{12}$, respectively.
Similarly, the elements of $Q_2$ commute with those of $Q_8$ and $Q_{11}$ and the elements of $Q_3$ with those of $Q_7$ and $Q_{10}$.
We also note that the elements of the Klein groups $K_i$ commute with the elements of a $Z_4$ subgroup:
the ones of $K_1$ commute with the ones of $Q_{10}$, the ones of $K_2$ with those of $Q_{12}$,  similarly the ones of $K_3$ with those of $Q_{11}$, 
the ones of $K_4$ commute with the ones of $Q_7$, the ones of $K_5$ with those of $Q_8$, as well as the ones of $K_6$ with those of $Q_9$.

Concerning the equivalence of the definition of the group $\Delta(96)$ in terms of the generators $S$ and $T$
and the definition in terms of $a$, $b$, $c$ and $d$, see eqs. (\ref{abcd1})-(\ref{abcd3}):
we can express the latter using $S$ and $T$ \cite{dATFH11}, namely
\be
\begin{array}{ll}
a=T^5ST^4~~,~~~~~~~~~~~~&b=ST^2ST^5\\
c=ST^2ST^4~~,~~~~~~~~~~~~&d=ST^2ST^6
\end{array}~~~,
\ee
and these fulfill the relations in eqs.(\ref{abcd1})-(\ref{abcd3}).

\mathversion{bold}
\subsection*{$\Delta(384)$}
\mathversion{normal}

In the case of $\Delta(384)$ various of the generating elements of the $Z_4$ and $Z_8$ subgroups $Q_i$ and $O_i$ commute, for example,
the elements of the $Z_4$ groups $Q_i$, $i=1,...,6$ commute. Furthermore, the elements of $Q_1$ commute with those contained in $Q_{15,16,17,18}$, respectively.
Similarly, the elements of $Q_2$ commute with the elements of the groups $Q_{8,10,12,14}$ and the elements of $Q_3$ commute with those of $Q_{7,9,11,13}$.
Among the $Z_8$ subgroups $O_i$ the elements of those with $i=1,...,12$ all commute. Apart from that the elements of $O_1$ commute with those of the groups
$O_{21,22,23,24}$, the elements of $O_2$ with the elements of $O_{15,16,19,20}$ and, finally, the elements of $O_3$ commute with those contained in $O_{13,14,17,18}$.
Also note that  $Q_i \subset O_i$ for $i=1,2,3$.

Since $\Delta(384)$ is a member of the series $\Delta(6 n^2)$ all statements made in the preceding subsections concerning the possible
definition of the group through the set of generators $a$, $b$, $c$ and $d$ also hold and a possible identification in the case of $\Delta(384)$ is  
\cite{dATFH11} 
\be
\begin{array}{ll}
a=T^{15}ST^8~~,~~~~~~~~~~~~&b=ST^6ST^3\\
c=ST^2ST^4~~,~~~~~~~~~~~~&d=ST^2ST^{14}
\end{array}~,
\ee
which satisfies eqs.(\ref{abcd1})-(\ref{abcd3}).

\newpage
\mathversion{bold}
\section{Tables of subgroups for $A_5$, $PSL(2,Z_7)$, $\Delta(96)$ and $\Delta(384)$}
\label{appA}
\mathversion{normal}

In the tables below all subgroups relevant for our discussion can be found together with their generating elements expressed in terms of $S$ and $T$.
\\[2cm]

\begin{table}[h]
\begin{center}
\begin{tabular}{|c|c|c||c|c|c|}
\hline
\multicolumn{2}{|c|}{{\tt Subgroup}} &
{\tt Generators} &
\multicolumn{2}{c|}{{\tt Subgroup}} &
{\tt Generators} \\
\hline
\multirow{5}{*}{$Z_2 \times Z_2$} & $K_1$ & $S$, $T^2 S T^3 S T^2$ &
\multirow{11}{*}{$Z_3$} & $C_1$ & $S T$ \\
& $K_2$ & $T^4 S T$, $S T^3 S T^2 S$ & & $C_2$ & $TS$ \\
& $K_3$ & $T S T^4$, $ S T^2 S T^3 S$ & & $C_3$ & $TST^3$ \\
& $K_4$ & $T^2 S T^3$, $S T^2 ST$ & & $C_4$ & $T^2 S T^2$ \\
& $K_5$ & $T^3 S T^2$, $T S T^2 S$ & & $C_5$ & $T^3 ST$ \\
\cline{1-3}
\multirow{6}{*}{$Z_5$} & $R_1$ & $T$ & &$C_6$ & $S T^3 ST$ \\
& $R_2$ & $ST^2$ & & $C_7$ & $S T^2 S T^3$ \\
& $R_3$ & $T^2 S$ & & $C_8$ & $ST^3 ST^2$ \\
& $R_4$ & $TST$ & & $C_9$ & $ST^2ST^4$ \\
& $R_5$ & $TST^2$ & & $C_{10}$ & $ST^2ST^2S$ \\
& $R_6$ & $T^2 ST$ & & & \\
\hline
\end{tabular}
\end{center}
\caption{\label{tab_abel_A5}List of generating elements of the subgroups of $A_5$.}
\end{table}
\vfill
\newpage
\begin{table}[h!]
\begin{center}
\begin{tabular}{|c|c|c||c|c|c|}
\hline
\multicolumn{2}{|c|}{{\tt Subgroup}} &
{\tt Generators} &
\multicolumn{2}{c|}{{\tt Subgroup}} &
{\tt Generators} \\
\hline
\multirow{14}{*}{$Z_2 \times Z_2$}
& $K_1$ & $S$, $T^2ST^3ST$              &
                                        \multirow{28}{*}{$Z_3$}
                                          & $C_1$ & $ST$ \\
& $K_2$ & $S$, $TST^3ST^2 $             & & $C_2$ & $TS$ \\
& $K_3$ & $T^4 ST^3 $, $T^2ST^4ST^2$    & & $C_3$ & $TST^5$ \\
& $K_4$ & $T^4ST^3$, $ST^4ST^3S$        & & $C_4$ & $T^2 S T^4$ \\
& $K_5$ & $T^5 ST^2$, $ST^4ST^3S$       & & $C_5$ & $T^3 ST^3$ \\
& $K_6$ & $T^2 ST^5$, $ST^3ST^4S$       & & $C_6$ & $T^4 ST^2$ \\
& $K_7$ & $T^3ST^4$, $ST^3ST^4S$        & & $C_7$ & $T^5ST$ \\
& $K_8$ & $T^3 ST^4$, $T^2 ST^4ST^2$    & & $C_8$ & $T ST^3S$ \\
& $K_9$ & $T^6ST $, $ST^5ST^6$          & & $C_9$ & $T^2 ST^4S$ \\
& $K_{10}$ & $TST^6$, $ST^4ST^4$        & & $C_{10}$ & $ST^2 ST^5$ \\
& $K_{11}$ & $T^2ST^5$, $ST^4ST^4$      & & $C_{11}$ & $ST^2 ST^4$ \\
& $K_{12}$ & $T^6ST$, $ST^3 ST^3$       & & $C_{12}$ & $T^4 ST^2S$ \\
& $K_{13}$ & $TST^6$, $ST^2ST$          & & $C_{13}$ & $ST^5 ST^2$ \\
& $K_{14}$ & $T^5 ST^2$, $ST^2ST$       & & $C_{14}$ & $ST^4 ST^2$ \\
& &                                     & & $C_{15}$ & $ST^3 ST$ \\
\cline{1-3}
\multirow{21}{*}{$Z_4$}
& $Q_1$ & $T^3S$                        & & $C_{16}$ & $ST^2 ST^4S$ \\ 
& $Q_2$ & $ST^3$                        & & $C_{17}$ & $ST^4 ST^2S$ \\
& $Q_3$ & $TST^3$                       & & $C_{18}$ & $TST^4 ST^5$ \\
& $Q_4$ & $T^2ST^2$                     & & $C_{19}$ & $TST^4 ST$ \\
& $Q_5$ & $TST^2$                       & & $C_{20}$ & $TST^3 ST^4$ \\
& $Q_6$ & $T^3ST$                       & & $C_{21}$ & $TST^5 ST^2$ \\
& $Q_7$ & $T^2ST$                       & & $C_{22}$ & $T^2ST^5 ST$ \\
& $Q_8$ & $TST^5S$                    & & $C_{23}$ & $TST^5 ST$ \\  
& $Q_9$ & $T^2ST^3S$                    & & $C_{24}$ & $ST^3 ST^5 ST$ \\
& $Q_{10}$ & $T^3ST^2S$                 & & $C_{25}$ & $ST^2 ST^4 ST^6$ \\
& $Q_{11}$ & $ST^3ST^4$                 & & $C_{26}$ & $ST^2 ST^4 ST^2$ \\
& $Q_{12}$ & $ST^4ST^3$                 & & $C_{27}$ & $ST^2 ST^4 ST^5$ \\
& $Q_{13}$ & $ST^2ST^3$                 & & $C_{28}$ & $ST^2 ST^4 ST$  \\
\cline{4-6}
& $Q_{14}$ & $ST^3ST^2$                 & \multirow{8}{*}{$Z_7$}  & $P_1$ & $T$    \\
& $Q_{15}$ & $ST^5ST$                   & & $P_2$ & $STS$   \\
& $Q_{16}$ & $ST^2ST^2S$                & & $P_3$ & $T^2S$   \\
& $Q_{17}$ & $TST^4ST^2$                & & $P_4$ & $TST^4$   \\
& $Q_{18}$ & $T^2ST^4ST$                & & $P_5$ & $T^4ST$   \\
& $Q_{19}$ & $TST^5ST^3$                & & $P_6$ & $ST^2$   \\
& $Q_{20}$ & $T^2ST^5ST^2$              & & $P_7$ & $T^2ST^3$   \\
& $Q_{21}$ & $T^3ST^5ST$                & & $P_8$ & $T^3ST^2$  \\
\hline
\end{tabular}
\end{center}
\caption{\label{tab_abel_PSL27}List of generating elements of the subgroups of $PSL(2,Z_7)$.}
\end{table}

\newpage
\begin{table}[h!]
\begin{center}
\begin{tabular}{|c|c|c||c|c|c|}
\hline
{\tt Subgroup}&&{\tt Generators}&{\tt Subgroup}&&{\tt Generators}\\
\hline
& $K$&$T^4,ST^4S,ST^4ST^4$& & &\\
\cdashline{2-3}
& $K_1$&$ST^4ST^4,S$& &$O_1$&$T$ \\
& $K_2$&$T^4,ST^2ST$& &$O_2$ &$ST^2$\\
$Z_2\times Z_2$& $K_3$&$ST^4S,T^7ST$&$Z_8$ &$O_3$ &$T^2S$\\
& $K_4$&$ST^4ST^4,T^6ST^2$& &$O_4$& $STS$\\
& $K_5$&$ST^4S,TST^7$& &$O_5$ &$ST^4ST$\\
& $K_6$&$T^4,ST^6ST^3$& &$O_6$ &$T^5ST$\\
\hline
& $C_1$&$ST$& & &\\
& $C_2$&$ST^3$& & &\\
& $C_3$&$ST^5$& &$Q_1$ &$T^2$\\
& $C_4$&$ST^7$& &$Q_2$ &$ST^2S$\\
& $C_5$&$T^2ST$& &$Q_3$ &$ST^2ST^2$\\
\cdashline{5-6}
& $C_6$&$T^2ST^3$& &$Q_4$ &$ST^4ST^2$\\
& $C_7$&$T^4ST$& &$Q_5$ &$ST^6ST^2$\\
$Z_3$& $C_8$&$T^3ST^4$&$Z_4$&$Q_6$&$ST^2ST^4$\\
\cdashline{5-6}
& $C_9$&$T^6ST$& &$Q_7$ &$ST^4$\\
& $C_{10}$&$TST^2$& &$Q_8$ &$T^3ST$\\
& $C_{11}$&$T^3ST^2$& &$Q_9$ &$ST^6ST$\\
& $C_{12}$&$T^5ST^2$& &$Q_{10}$ &$T^2ST^2$\\
& $C_{13}$&$T^4ST^3$& &$Q_{11}$ &$TST^3$\\
& $C_{14}$&$T^2ST^5$& &$Q_{12}$ &$ST^2ST^3$\\
& $C_{15}$&$TST^4$& & &\\
& $C_{16}$&$TST^6$& & &\\
\hline
\end{tabular}
\end{center}
\caption{List of generating elements of the subgroups  $Z_3$, $Z_4$, $Z_2 \times Z_2$ and $Z_8$ of $\Delta(96)$.
\label{tab_abel_D96}}
\end{table}

\begin{table}[h!]
\begin{center}
\begin{tabular}{|c|c||c|c||c|c||c|c||c|c|}
\hline
\multicolumn{10}{|c|}{{\tt $Z_{2}$ groups}}  \\
\hline
$V_1$ & $ST^6ST^3$ & $V_2$ & $ST^2ST$ & $V_3$ & $TST^2S$ & $V_4$ & $T^4ST^2ST$ & $V_5$ & $TST^7$\\
$V_6$ & $S$ & $V_7$ & $T^2ST^6$ & $V_8$ & $T^5ST^3$ & $V_9$ & $T^4ST^4$ & $V_{10}$ & $T^7ST$\\
$V_{11}$ & $T^3ST^5$ & $V_{12}$ & $T^6ST^2$ & $V_{13}$ & $ST^4S$ & $V_{14}$ & $ST^4ST^4$ & $V_{15}$ & $T^4$\\
\hline
\end{tabular}
\end{center}
\caption{\label{tab_Z2_D96} List of generating elements of the $Z_{2}$ subgroups of $\Delta(96)$.}
\end{table}
\newpage
\begin{table}[h!]
\begin{center}
\begin{tabular}{|c|c||c|c|}
\hline
\multicolumn{4}{|c|}{{\tt Klein groups}}  \\
\hline
$K$ & $T^8$, $ST^8S$, $ST^8ST^8$ & $K_1$ & $ST^8S$, $T^{11}ST^5$ \\
\cdashline{1-2}
$K_2$ & $ST^8S$, $TST^{15}$ & $K_3$ & $ST^8S$, $T^{15}ST$ \\
$K_4$ & $ST^8S$, $T^5ST^{11}$ & $K_5$ & $ST^8ST^8$, $T^6ST^{10}$ \\
$K_6$ & $ST^8ST^8$, $T^4ST^{12}$ & $K_7$ & $ST^8ST^8$, $T^2ST^{14}$ \\
$K_8$ & $ST^8ST^8$, $S$ & $K_9$ & $T^8$, $ST^{14}ST^7$ \\
$K_{10}$ & $T^8$, $ST^{10}ST^5$ &  $K_{11}$ & $T^8$, $ST^6ST^3$\\ 
$K_{12}$ & $T^8$, $ST^2ST$ && \\
\hline
\end{tabular}
\end{center}
\caption{\label{tab_Klein_D384} List of generating elements of the Klein groups contained in $\Delta(384)$.}
\end{table}

\begin{table}[h!]
\begin{center}
\begin{tabular}{|c|c||c|c||c|c||c|c|}
\hline
\multicolumn{8}{|c|}{{\tt $Z_3$ groups}}  \\
\hline
$C_1$ & $ST$ & $C_2$ & $TS$ & $C_3$ & $ST^3$ & $C_4$ & $T^3S$ \\
$C_5$ & $T^5S$ & $C_6$ & $T^7S$ & $C_7$ & $T^9S$ & $C_8$ & $T^{11}S$ \\
$C_9$ & $TST^{14}$ & $C_{10}$ & $T^2ST^{13}$ & $C_{11}$ & $T^3ST^{12}$ & $C_{12}$ & $T^4ST^{11}$\\
$C_{13}$ & $T^5ST^{10}$ & $C_{14}$ & $T^6ST^9$ & $C_{15}$ & $T^7ST^8$ & $C_{16}$ & $T^8ST^7$ \\
$C_{17}$& $T^9ST^6$ & $C_{18}$ & $T^{10}ST^5$ & $C_{19}$ & $T^{11}ST^4$ & $C_{20}$ & $T^{12}ST^3$ \\
$C_{21}$ & $T^{13}ST^2$ & $C_{22}$ & $T^{14}ST$ & $C_{23}$ & $TST^{12}$ & $C_{24}$ & $T^2ST^{11}$ \\
$C_{25}$ & $T^3ST^{10}$ & $C_{26}$ & $T^4ST^9$ & $C_{27}$ & $T^5ST^8$ & $C_{28}$ & $T^6ST^7$ \\
$C_{29}$ & $T^7ST^6$ & $C_{30}$ & $T^8ST^5$ & $C_{31}$ & $T^9ST^4$ & $C_{32}$ & $T^{10}ST^3$ \\
$C_{33}$ & $T^{11}ST^2$ & $C_{34}$ & $T^{12}ST$ & $C_{35}$ & $TST^{10}$ & $C_{36}$ & $T^2ST^9$ \\
$C_{37}$ & $T^3ST^8$ & $C_{38}$ & $T^4ST^7$ & $C_{39}$ & $T^5ST^6$ & $C_{40}$ & $T^6ST^5$ \\
$C_{41}$ & $T^7ST^4$ & $C_{42}$ & $T^8ST^3$ & $C_{43}$ & $T^9ST^2$ & $C_{44}$ & $T^{10}ST$ \\
$C_{45}$ & $TST^8$ & $C_{46}$ & $T^2ST^7$ & $C_{47}$ & $T^3ST^6$ & $C_{48}$ & $T^4ST^5$ \\
$C_{49}$ & $T^5ST^4$ & $C_{50}$ & $T^6ST^3$ & $C_{51}$ & $T^7ST^2$ & $C_{52}$ & $T^8ST$ \\
$C_{53}$ & $TST^6$ & $C_{54}$ & $T^2ST^5$ & $C_{55}$ & $T^3ST^4$ & $C_{56}$ & $T^4ST^3$ \\
$C_{57}$ & $T^5ST^2$ & $C_{58}$ & $T^6ST$ & $C_{59}$ & $TST^4$ & $C_{60}$ & $T^2ST^3$ \\
$C_{61}$ & $T^3ST^2$ & $C_{62}$ & $T^4ST$ & $C_{63}$ & $TST^2$ & $C_{64}$ & $T^2ST$ \\ 
\hline
\end{tabular}
\end{center}
\caption{\label{tab_Z3_D384} List of generating elements of the $Z_3$ subgroups of $\Delta(384)$.}
\end{table}
\begin{table}[h!]
\begin{center}
\begin{tabular}{|c|c||c|c||c|c||c|c|}
\hline
\multicolumn{8}{|c|}{{\tt $Z_4$ groups}}  \\
\hline
$Q_1$ & $T^4$ & $Q_2$ & $ST^4S$ & $Q_3$ & $ST^4ST^4$ & & \\
\cdashline{1-6}
$Q_4$ & $ST^4ST^8$ & $Q_5$ & $ST^8ST^4$ & $Q_6$ & $ST^4ST^{12}$ & & \\
\cdashline{1-6}
$Q_7$ & $ST^8$ & $Q_8$ & $TST^7$ & $Q_9$ & $T^2ST^6$ & $Q_{10}$ & $T^3ST^5$ \\
$Q_{11}$ & $T^4ST^4$ & $Q_{12}$ & $T^5ST^3$ & $Q_{13}$ & $T^6ST^2$ & $Q_{14}$ & $T^7ST$ \\
$Q_{15}$ & $ST^2ST^5$ & $Q_{16}$ & $ST^6ST^7$ & $Q_{17}$ & $ST^{10}ST^9$ & $Q_{18}$ & $ST^{14}ST^{11}$ \\
\hline
\end{tabular}
\end{center}
\caption{\label{tab_Z4_D384} List of generating elements of the $Z_4$ subgroups of $\Delta(384)$.}
\end{table}

\begin{table}[h!]
\begin{center}
\begin{tabular}{|c|c||c|c||c|c|}
\hline
\multicolumn{6}{|c|}{{\tt $Z_8$ groups}}  \\
\hline
$O_1$ & $T^2$ & $O_2$ & $ST^2S$ & $O_3$ & $ST^{14}ST^{14}$  \\
\hdashline
$O_4$ & $ST^6ST^4$ & $O_5$ & $ST^4ST^6$ & $O_6$ & $ST^{10}ST^{14}$ \\
\hdashline
$O_7$ & $ST^2ST^4$ & $O_8$ & $ST^2ST^{14}$ & $O_9$ & $ST^4ST^2$  \\
\hdashline
$O_{10}$ & $ST^2ST^8$ & $O_{11}$ & $ST^6ST^{14}$ & $O_{12}$ & $ST^8ST^2$ \\
\hdashline
$O_{13}$ & $ST^4$ & $O_{14}$ & $T^4S$ & $O_{15}$ & $TST^{11}$ \\
$O_{16}$ & $TST^3$ & $O_{17}$ & $T^2ST^{10}$ & $O_{18}$ & $T^2ST^2$ \\
$O_{19}$ & $T^3ST^9$ & $O_{20}$ & $T^3ST$ & $O_{21}$ & $ST^2ST^3$ \\
$O_{22}$ & $ST^6ST$ & $O_{23}$ & $ST^{10}ST^3$ & $O_{24}$ & $ST^{14}ST$\\
\hline
\end{tabular}
\end{center}
\caption{\label{tab_Z8_D384} List of generating elements of the $Z_8$ subgroups of $\Delta(384)$.}
\end{table}

\begin{table}[h!]
\begin{center}
\begin{tabular}{|c|c||c|c||c|c||c|c|}
\hline
\multicolumn{8}{|c|}{{\tt $Z_{16}$ groups}}  \\
\hline
$Y_1$ & $T$ & $Y_2$ & $ST^2$ & $Y_3$ & $STS$ & $Y_4$ & $ST^6$ \\
$Y_5$ & $ST^{10}$ & $Y_6$ & $ST^{14}$ & $Y_7$ & $TST^5$ & $Y_8$ & $TST^9$ \\
$Y_9$ & $TST^{13}$ & $Y_{10}$ & $ST^4ST$ & $Y_{11}$ & $ST^8ST$ & $Y_{12}$ & $ST^{12}ST$ \\
\hline
\end{tabular}
\end{center}
\caption{\label{tab_Z16_D384} List of generating elements of the $Z_{16}$ subgroups of $\Delta(384)$.}
\end{table}

\begin{table}[h!]
\begin{center}
\begin{tabular}{|c|c||c|c||c|c||c|c|}
\hline
\multicolumn{8}{|c|}{{\tt $Z_{2}$ groups}}  \\
\hline
$V_1$ & $TST^2S$ & $V_2$ & $ST^{14}ST^7$ & $V_3$ & $T^3ST^6S$ & $V_4$ & $T^5ST^{10}S$ \\
$V_5$ & $ST^2ST^9$ & $V_6$ & $ST^6ST^3$ & $V_7$ & $ST^{10}ST^5$ & $V_8$ & $ST^2ST$ \\
$V_9$ & $TST^{15}$ & $V_{10}$ & $ST^{13}ST^{10}$ & $V_{11}$ & $T^5ST^{11}$ & $V_{12}$ & $TST^3ST^5$ \\
$V_{13}$ & $T^{15}ST$ & $V_{14}$ & $T^4ST^{12}$ & $V_{15}$ & $T^6ST^{10}$ & $V_{16}$ & $T^8ST^8$ \\
$V_{17}$ & $ST^6ST^{10}S$ & $V_{18}$ & $ST^3ST^6$ & $V_{19}$ & $ST^8ST^{12}ST^{12}$ & $V_{20}$ & $T^2ST^{14}$ \\
$V_{21}$ & $T^{11}ST^5$ & $V_{22}$ & $S$ & $V_{23}$ & $ST^9ST^2$ & $V_{24}$ & $T^9ST^7$ \\
$V_{25}$ & $ST^8S$ & $V_{26}$ & $ST^8ST^8$ & $V_{27}$ & $T^8$ &  & \\
\hline
\end{tabular}
\end{center}
\caption{\label{tab_Z2_D384} List of generating elements of the $Z_{2}$ subgroups of $\Delta(384)$.}
\end{table}
$\phantom{xxx}$\\[2in]
$\phantom{xxx}$\\[2in]


\end{document}